%% file: paper.tex
\documentclass[useAMS,usenatbib]{mn2e}

\usepackage{graphicx}
\usepackage{hyperref}
\usepackage{multirow}
\usepackage{lscape}
\usepackage{upgreek}

\newcommand{\haslam}{sync[H0.4]} 
\newcommand{\jonas}{sync[J2.3]} 
\newcommand{\ddd}{\ha[DDD]} 
\newcommand{\dddfd}{\ha[DDD0.33]} 
\newcommand{\fink}{\ha[F03]} 
\newcommand{\fds}{dust[FDS94]}
\newcommand{\ha}{H$\alpha$} 

\def\Planck{{\it Planck\/}}
\def\WMAP{{\it WMAP\/}}
\def\IRAS{{\it IRAS\/}}
\def\DIRBE{{\it COBE}-DIRBE\/}

\def\pdeg{\ifmmode $\setbox0=\hbox{$^{\circ}$}\rlap{\hskip.11\wd0 .}$^{\circ}
          \else \setbox0=\hbox{$^{\circ}$}\rlap{\hskip.11\wd0 .}$^{\circ}$\fi}

\usepackage{color}
\newcommand{\ch}[1]{#1}

\title{Template fitting of \WMAP~7-year data: anomalous dust or flattening synchrotron emission?}
\author[M.~W.~Peel et al.]
 {M.\,W.\,Peel$^1$, C.\,Dickinson$^1$, R.\,D.\,Davies$^1$, A.\,J.\,Banday$^{2,3}$, T.\,R.\,Jaffe$^{2,3}$, J.\,L.\,Jonas$^4$\\
 $^1$ Jodrell Bank Centre for Astrophysics, Alan Turing Building, School of Physics and Astronomy, The University of Manchester,\\
\phantom{$^1$} Oxford Road, Manchester, M13 9PL, UK\\
 $^2$ UniversitŽ de Toulouse, UPS-OMP, IRAP, F-31028 Toulouse, France\\
$^3$ CNRS, IRAP, 9 Av. colonel Roche, BP 44346, F-31028 Toulouse cedex 4, France\\
$^4$ Rhodes University, Department of Physics and Electronics, Grahamstown 6140, South Africa}
\begin{document}

\pagerange{\pageref{firstpage}--\pageref{lastpage}} \pubyear{2011}
\date{Accepted 2012 May 21.  Received 2012 April 23; in original form 2011 December 5}
\pubyear{2012}
\maketitle

\label{firstpage}

\begin{abstract}
Anomalous microwave emission at 20-40\,GHz has been detected across our Galactic sky. It is highly correlated with thermal dust emission and hence it is thought to be due to spinning dust grains. Alternatively, this emission could be due to synchrotron radiation with a flattening (hard) spectral index. We cross-correlate synchrotron, free-free and thermal dust templates with the \WMAP~7-year maps using synchrotron templates at both 408\,MHz and 2.3\,GHz to assess the amount of flat synchrotron emission that is present, and the impact that this has on the correlations with the other components. We find that there is only a small amount of flattening visible in the synchrotron spectral indices by 2.3\,GHz, of around $\Delta \beta \approx 0.05$, and that the significant level of dust-correlated emission in the lowest \WMAP~bands is largely unaffected by the choice of synchrotron template, particularly at high latitudes (it decreases by only $\sim$7 per cent when using 2.3\,GHz rather than 408\,MHz). This agrees with expectation if the bulk of the anomalous emission is generated by spinning dust grains.
\end{abstract}

\begin{keywords}
diffuse radiation -- Galaxy: general -- radiation mechanisms: general -- radio continuum: ISM
\end{keywords}

\section{Introduction} \label{sec:introduction}
Maps of high-frequency ($>10$\,GHz) radio emission contain multiple distinct astrophysical components that must be disentangled to study both the astrophysics of our Galaxy and the cosmological information contained in the Cosmic Microwave Background (CMB) radiation. In particular, multifrequency microwave instruments such as the Wilkinson Microwave Anisotropy Probe \citep[{\it WMAP},][]{2003Bennettb} and the \Planck~satellite \citep{2010Tauber} primarily measure combinations of synchrotron emission from electrons spiralling around magnetic fields (low frequencies); free-free emission from the interaction of electrons with ions (intermediate frequencies); anomalous microwave emission (AME) thought to be from spinning dust grains (peaking at $\sim$20-40\,GHz; see e.g. \citealp{1998Draine,2004deOliveiraCosta,2005Watson,2010Ysard,2011PlanckAnom}); CMB emission (intermediate to high frequencies); and thermal dust emission (high frequencies).

These various components can be separated from each other by fitting templates of the different emission mechanisms to the data, i.e. using maps of single components to separate the high frequency maps based on the emission morphology (see e.g. \citealp{1996Kogut,2003Banday,2006Davies}). However, problems arise when the morphology of the emission changes with frequency, which can result from the differing spectral indices of the various astrophysical objects that are generating that emission. In particular, there is a wide range of observed synchrotron spectral indices, and the overall behaviour of the emission may change at higher frequencies -- either steepening due to synchrotron ageing, or flattening due to a recent injection of energetic electrons or multiple components of emission.

\begin{figure}
\centering
\includegraphics[scale=0.34]{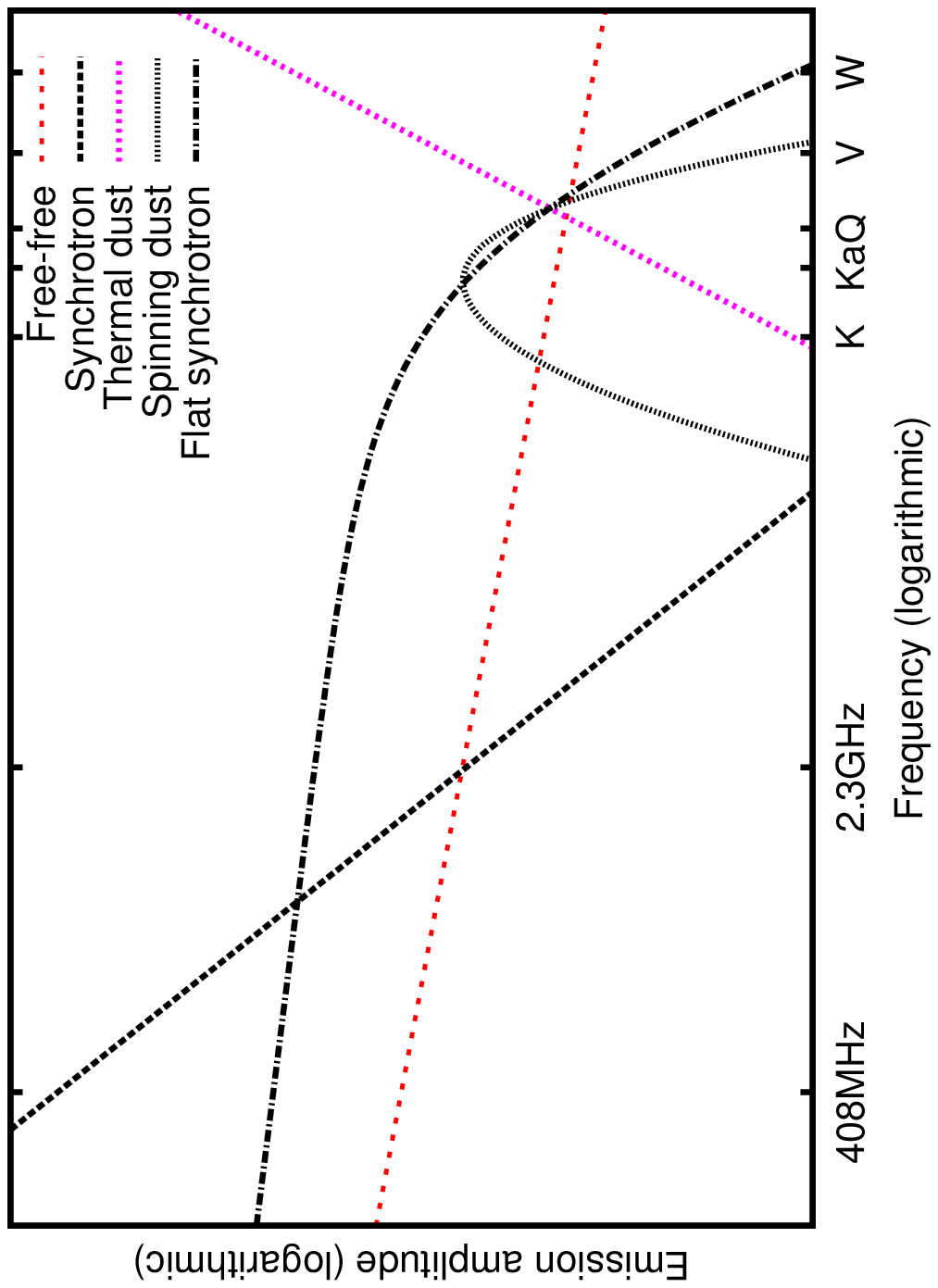}
\caption{\ch{A sketch indicating the different emission mechanisms present at various frequencies. Spectral shapes are plotted in arbitrary flux density units, with spectral indices $\alpha=\beta+2$. Two possible synchrotron components are shown in black: one steep spectrum ($S\propto\nu^{-1.0}$), the other flat spectrum but turning over at WMAP frequencies ($S\propto\nu^{-0.1}/[1+[\nu/45\,\mathrm{GHz}]^{2}]$). An illustrative spinning dust model is also shown in grey. Free-free emission ($S\propto\nu^{-0.15}$) is shown in red, and thermal dust ($S\propto\nu^{1.5}$) in magenta. The cause of the anomalous emission at WMAP frequencies could be due to either spinning dust or flat synchrotron emission that turns over at WMAP frequencies.}}
\label{fig:schematic}
\end{figure}

In this context, \citet{2003Bennett} have suggested that the dust-correlated emission observed by \WMAP~could be due to flat (hard) synchrotron radiation (with a spectral index $\beta\sim-2.5$) that is not well traced by the standard low-frequency map of synchrotron emission, as opposed to being due to a different emission mechanism such as spinning dust. These two scenarios are \ch{sketched out} in Figure \ref{fig:schematic}. In this paper, we apply the template fitting method to the \WMAP~7-year maps using synchrotron templates at both 408\,MHz and 2.3\,GHz. The 408\,MHz map is the standard tracer of synchrotron radiation used in this type of analysis, however due to its low frequency it will predominantly trace steep spectrum radiation, whereas the 2.3\,GHz map at nearly 6 times higher frequency will trace more of flatter-spectrum radiation. We look for an increased correlation between the \WMAP~maps and the higher frequency synchrotron template, and a corresponding reduced correlation with the dust template, which would be indicative of the anomalous emission being flat spectrum synchrotron radiation.

In order to judge how sensitive the 2.3\,GHz map might be to flat-spectrum synchrotron compared with the 408\,MHz map, we use the r.m.s. values of the 408\,MHz and FDS maps (5.9\,K and 6.8\,$\upmu$K respectively\ch{,} see \citealp{2006Davies}) combined with the template coefficients measured later in this paper, and assume spectral indices of $\beta=-3.01$ for steep-spectrum emission and $-2.5$ for the flat spectrum emission. Assuming that all of the dust-correlated emission measured at 22.8\,GHz is flat spectrum emission, and assuming no spectral curvature in either the steep- or flat-spectrum component, then it would constitute 29 per cent of the 2.3\,GHz emission, compared with 14 per cent of the 0.408\,GHz emission. Thus, we would expect to see a considerable difference in the template coefficients. Given this, one might expect the 22.8\,GHz \ch{dust template coefficient} to change by a factor $\sim14/29$ if the dust-correlated component were completely due to a flatter spectrum synchrotron spectrum. Though there is evidence of curvature in the synchrotron spectrum, these figures give a qualitative idea of the difference in relative strengths of the components in the low-frequency templates that motivates this work.
% Steep-spectrum Synchrotron:
% 5.9*6.8*(22.8/0.408)^3.01 = 7.285K
% 5.9*6.8*(22.8/2.3)^3.01 = 0.040K
% Flat-spectrum:
% 6.8*7.9 *(22.8/0.408)^(2.5) = 1.25K -> 14% (1.25/(1.25+7.285))
% 6.8*7.9 *(22.8/2.3)^(2.5) = 0.017K -> 29% (17/(17+40))
% 6.8*7.9 *(22.8/0.408)^(2.85) = 5.13K -> 41%
% 6.8*7.9 *(22.8/2.3)^(2.85) = 0.037K -> 48%

In Section \ref{sec:maps} we discuss the data used here, and the choice of the specific foreground templates. We briefly summarise the cross-correlation method that we use to carry out the template fitting in Section \ref{sec:correlation}. Our results are presented in Section \ref{sec:results}, along with an assessment of their robustness as a function of resolution, Galactic masks, and different template options. \ch{This section also} looks at a few selected areas of Galactic emission, where we confirm that our results are valid both for the whole sky and for sections of it. We conclude in Section \ref{sec:conclusion}.

\section{Data and templates} \label{sec:maps}
For this analysis, we use a combination of low frequency maps from ground-based radio telescopes which map the synchrotron radiation (described in section 2.1); optical \ha~emission maps tracing the free-free emission (section 2.2) and frequency-scaled maps of the dust emission from \IRAS~(section 2.3) as our foreground templates, and which we then fit to the maps from \WMAP~(section 2.4). The properties of the maps used here are summarised in table \ref{tab:input_maps}.

All of the maps are in the {\sc Healpix}\footnote{\ch{\url{http://healpix.jpl.nasa.gov}}} pixelization scheme \citep{2005Gorski}. The maps are smoothed to the desired resolution (nominally 3$^\circ$) whilst in their original $N_\mathrm{side}$ (either 512 or 256) before being down-sampled to a nominal $N_\mathrm{side}=64$ (i.e. a pixel size of 55 arcmin). We mask all of the maps using the extent of the 2.3\,GHz \ch{observations. In} addition, we  use a combination of \WMAP~Galactic plane and point source masks to exclude strong emission that may bias the results (particularly due to the absorption of \ha~by thermal dust).

Throughout this work we use the spectral index convention of $T \propto \nu^\beta$. We use Rayleigh-Jeans (RJ) antenna temperatures \ch{throughout. Where} maps are in CMB temperatures we convert to antenna temperatures using $T_\mathrm{RJ} = T_\mathrm{CMB} \phantom{.} x^2 e^x / (e^x-1)^2$, where $x = h \nu / k T_\mathrm{CMB}$, in which $T_\mathrm{CMB}=2.725$\ch{\,K}.

\subsection{Synchrotron templates}
The standard all-sky synchrotron map, which is used in a wide variety of component separation analyses, is the 408\,MHz map \citep{1981Haslam,1982Haslam}, hereafter \haslam. This is currently the only high resolution ($\sim1^\circ$) map of the entire sky that is available at a frequency of less than a \ch{gigahertz}. It combines observations made with four large single-dish telescopes (Jodrell Bank Mk1 and Mk1A, Effelsberg and Parkes) and is absolutely calibrated using the 404\,MHz survey by \citet{1962Pauliny-Toth} (extrapolated between 404 and 408\,MHz using a spectral index of $-2.5$). The 408\,MHz map hence has an accuracy of $\sim3$\,K in absolute temperature and $\pm$10 per cent in $T_\mathrm{b}$ scale.

There are a number of different versions of this available (see e.g. \citealp{1996Davies,2003Platania})\ch{. We} use the version available from the Legacy Archive for Microwave Background Data Analysis (LAMBDA)\footnote{\url{http://lambda.gsfc.nasa.gov/product/foreground/haslam\_408.cfm}}. This version has been de-striped and point sources have been removed. Although the resolution of the survey is nominally 0\pdeg85, the post-processing that has been applied to it has smoothed the map slightly; we assume that the map now has 1$^\circ$ resolution. As we are looking at the diffuse emission, which is not significantly affected by the precise beam size, this assumption is sufficient for our analysis.

At $5.7$ times higher frequency than \haslam, the 2.326\,GHz map of \citet{1998Jonas}, hereafter \jonas, made from observations with the 26-m HartRAO telescope, can also be used as a synchrotron template. Although not a full sky map, this still covers 67 per cent of the sky at high resolution (20 arcmin) and high sensitivity. The map was made differentially, with an arbitrary zero point that was determined using the 2.0\,GHz absolute measurements by \citet{1994Bersanelli} (assuming a spectral index between 2.0 and 2.3\,GHz of $-2.75$), and an uncertainty in $T_\mathrm{b}$ of $\sim$5 per cent. As such, the map should be completely independent of \haslam, and in the presence of flat-spectrum synchrotron emission it should be a better tracer of the high-frequency synchrotron that is seen by \WMAP. 

\begin{figure*}
\centering
\includegraphics[scale=0.3]{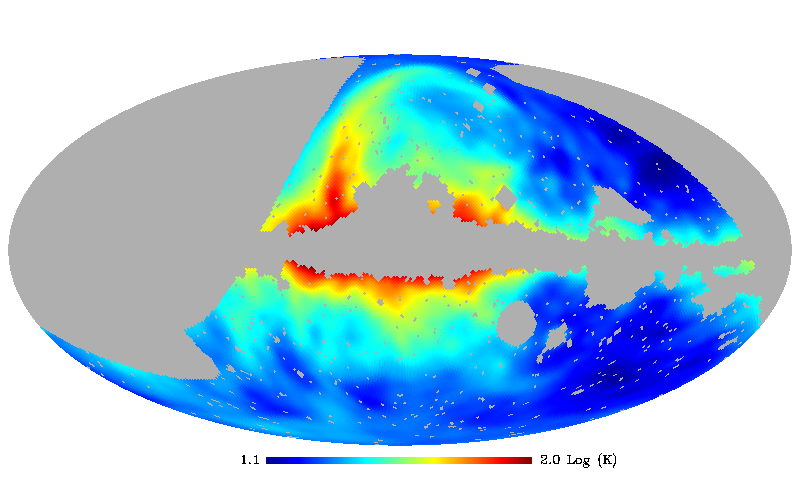}
\includegraphics[scale=0.3]{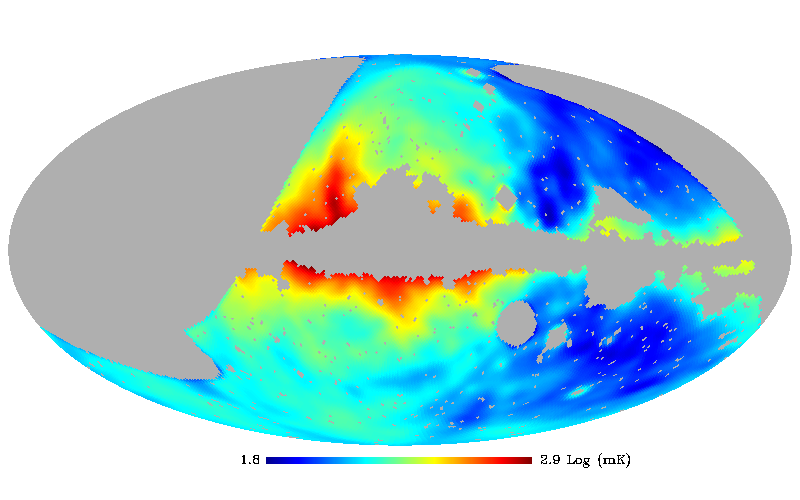}\
\includegraphics[scale=0.3]{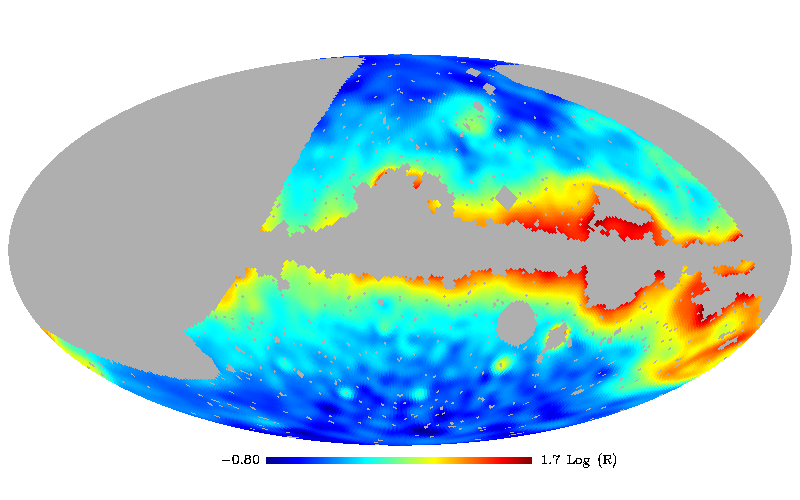}
\includegraphics[scale=0.3]{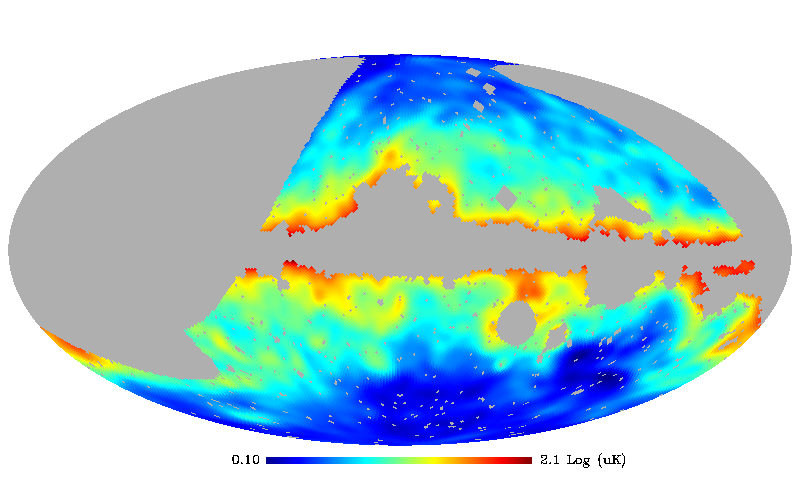}
\caption{The input templates, smoothed to 3$^\circ$, using $N_\mathrm{side}=64$ and plotted with a \ch{logarithmic} colour scheme: \haslam~(top-left), \jonas~(top-right), \ddd~(bottom-left) and \fds~(bottom-right). \ch{The maps are in Mollweide projection and are in Galactic coordinates with longitude increasing to the left.} All maps have been masked with the 2.3\,GHz survey area and the KQ85 mask. The morphological differences between the different emission mechanisms can be clearly seen.}
\label{fig:templates}
\end{figure*}

The two synchrotron maps are shown in Figure \ref{fig:templates}. The change in the morphology of the synchrotron emission between these two frequencies is \ch{obvious. For} example, the north polar spur has a steep synchrotron spectrum and as such it is much less noticeable in \jonas~than \haslam.

We note that a 1.4\,GHz map also exists \citep{1982Reich,1986Reich,2001Reich}, however we do not make use of this here as the higher frequency map will provide a better tracer of flat spectrum synchrotron radiation. In the future, this work can be repeated using a 5\,GHz map from C-BASS \citep{2010King}, which will provide an even better template of flat spectrum synchrotron radiation.

\begin{table}
\caption{The properties of the template maps (top) and input maps (bottom) used in this analysis. Where maps have been pre-smoothed, we provide both the smoothed and raw resolutions. \ch{H$\alpha$ is used as a proxy for free-free emission, so we do not quote a frequency for it here.} The thermal noise per observation, $\sigma$, is relevant for the \WMAP~maps; the others are systematics-dominated.}
    \tabcolsep 4.5pt
    \small
\begin{tabular}{lcccc}
\hline
Name & \ch{Freq. (GHz)} & Res. & \ch{$\sigma_0$} (RJ) & Notes\\
\hline
\haslam & 0.408 & 60\arcmin & -- & orig. 51\arcmin\\
\jonas & 2.326 & 60\arcmin & -- & orig. 21\arcmin\\
\fds & 94 & 6\arcmin & -- & \\
\fink & -- & $\sim$6\arcmin & -- & Mixed res.\\
\ddd & -- & 60\arcmin & -- & $f_{d}=0$\\
\dddfd & -- & 60\arcmin & -- & $f_{d}=0.33$\\
\hline
\WMAP~K & 22.8 & 60\arcmin & 1.418~mK & orig. 0\pdeg88\\ %Noise orig 1.437 mKCMB
\WMAP~Ka & 33.0 & 60\arcmin & 1.429~mK & orig. 0\pdeg66\\ % 1.470 mKCMB
\WMAP~Q & 40.7 & 60\arcmin & 2.105~mK & orig. 0\pdeg51\\ % 2.197 mKCMB
\WMAP~V & 60.8 & 60\arcmin & 2.854~mK & orig. 0\pdeg35\\ % 3.137 mKCMB
\WMAP~W & 93.5 & 60\arcmin & 5.253~mK & orig. 0\pdeg22\\ % 6.549 mKCMB
\hline
\end{tabular}
\label{tab:input_maps}
\end{table}

\subsection{Free-free templates}
We use a map of the \ha~line emission at 656.28~nm in units of Reyleigh (R) as a proxy for the free-free emission, as is conventional for component separation due to the difficulty in cleanly separating large-scale free-free from synchrotron emission at radio frequencies. There are, however, a number of issues with any map of \ha~\ch{emission} -- including geocoronal emission from the Earth's upper atmosphere, baseline effects, stellar residuals and dust absorption.  There are two full-sky \ha~maps available, both of which attempt to mitigate these \ch{effects, namely} those by \citet{2003Dickinson} and \citet{2003Finkbeiner}. Both of these are composites of several different large-scale ground-based surveys.

The map produced by \citet{2003Finkbeiner}, hereafter \fink, combines data from the Wisconsin H-alpha Mapper survey \citep[WHAM;][]{1998Reynolds}, the Southern H Alpha Sky Survey Atlas \citep[SHASSA;][]{2001Gaustad} and Virginia Tech Spectral line Survey \citep[VTSS;][]{1998Dennison}. Unfortunately, these three surveys have different resolutions and fields of view, which means that the combined map cannot be easily smoothed to a single resolution. We assume that the resulting map has a resolution of 6 arcmin, which will lead to over-smoothing in parts of the map where the observations were made at a lower resolution (see \citealp{2011Ghosh} for a power-spectrum analysis of this issue). Again, as we are focusing on the diffuse emission, this should not have a significant effect on our results. Additionally, most of the fields where this is the case are in the Northern hemisphere; of the Southern area that we are using, only 10 per cent of the \jonas~survey area is affected (or 7 per cent when the KQ85 mask is applied).

The maps produced by \citet[][]{2003Dickinson} avoid this resolution issue by combining only WHAM and SHASSA with suitable smoothing to a resolution of 1$^\circ$. We use two versions of this map -- one map that has been corrected for dust absorption of the \ha~emission using the 100~$\upmu$m dust map of \citet{1998Schlegel} with an effective fraction of the dust that contributes towards \ha~absorption of $f_\mathrm{d}=0.33$ (\dddfd), and a second map without this correction factor (\ddd) for more direct comparison with the results using \fink.

We use the \ddd~map as our standard map, and will investigate the effects of using the \fink~and \dddfd~maps in section 4.2. We correct the synchrotron maps for free-free contamination using the \ha~maps. We use the values presented in \citet{2003Dickinson} assuming that the gas has an electron temperature $T_\mathrm{e}=7000$\,K everywhere on the sky (although the electron temperature is known to vary across the sky, this has yet to be quantified and mapped). At 408\,MHz this correction factor is 51.2~mK/R and at 2.326\,GHz it is 1.33~mK/R. The inclusion of these free-free corrections to the synchrotron templates does not significantly alter our later results.

\subsection{Thermal dust}
We use the thermal dust map created by \citet[][hereafter \fds]{1999Finkbeiner} via an extrapolation of the 100\,$\upmu$m {\it IRAS} and 100/240\,$\upmu$m \DIRBE~maps to 94\,GHz. This map has a resolution of 6~arcmin. We use the version available from LAMBDA\footnote{\url{http://lambda.gsfc.nasa.gov/product/foreground/dust_map.cfm}}. In the future, maps of the cold thermal dust as seen by \Planck~HFI at 857 or 545\,GHz can be used either directly as dust templates, or as extra information in a modified extrapolation of the dust to low frequencies, which will better trace the thermal dust seen at frequencies $\sim100$\,GHz.

\subsection{\WMAP~7-year data}\label{sec:wmap7}
We use the 1$^\circ$-smoothed \WMAP~7-year maps \citep{2011Jarosik} provided by LAMBDA\footnote{\url{http://lambda.gsfc.nasa.gov/product/map/dr4/skymap_info.cfm}}, which we then smooth further to the required resolution. This approach avoids problems with e.g. non-circular beams, which will have been taken into account by \WMAP. The maps are at frequencies of 22.8, 33.0, 40.7, 60.8 and 93.5\,GHz.

We also require the number of observations \ch{per pixel, $N_\mathrm{obs}$, for the \WMAP~maps after they have been smoothed and their pixel resolution has been degraded}. The effective $N_\mathrm{obs}$ maps were calculated using 1000 Monte-Carlo simulations. Each noise realization, based on a Gaussian white noise with $\sigma=\sigma_0 / \sqrt{N_\mathrm{obs}}$\ch{ where $\sigma_0$ is the nominal thermal noise per observation (see Table 1)}, was smoothed to the appropriate resolution and degraded to the final $N_\mathrm{side}$. The effective $N_\mathrm{obs}$ was then calculated using the r.m.s. scatter of the realizations for each pixel to get the final noise level, and renormalizing to obtain the effective $N_\mathrm{obs}$ for each pixel \ch{for the maps as used in the analysis}.

We mask the Galactic plane using three different masks: the \WMAP~7-year extended temperature mask, ``KQ75'', which masks 29.4 per cent of the sky; the \WMAP~7-year temperature mask, ``KQ85'', which masks out 21.7 per cent of the sky \citep{2010Gold}; and the ``kp2'' mask from the \WMAP~1-year release, which masks only 15 per cent of the sky. Using these masks excludes areas of the sky that are very bright in emission that could bias the results, particularly for fitting the free-free emission since \ha~along the Galactic plane is absorbed by dust. In addition to masking the galactic plane, all of these masks also exclude the pixels that contain bright extragalactic and other compact \ch{sources. Although} there are some differences between the 1 and 7 year point source masks, this will be small since these sources will be faint, and are further diluted by the map smoothing applied here.

\section{Cross-correlation method} \label{sec:correlation}
\ch{We cross-correlate each \WMAP~map with the emission templates by carrying out a least-squares fit to minimise the $\chi^2$, via a direct matrix inversion, as per  \citet{2006Davies} and \citet{2011Ghosh}. The cross--correlation measure, $\alpha$, between a data vector, ${\bf d}$ and a template vector ${\bf t}$ can be measured by
minimising:
\begin{equation}
\chi^2 = ({\bf d}-\alpha {\bf t})^T \cdot {\bf M}^{-1}_{SN}\cdot ({\bf d}-\alpha {\bf t}) = {\bf \tilde{d}}^T \cdot {\bf M}^{-1}_{SN}\cdot {\bf \tilde{d}}
\end{equation}
where ${\bf M}_{SN}$ is the covariance matrix including both signal and noise for the template--corrected data vector ${\bf \tilde{d}}
\equiv {\bf d} - \alpha {\bf t}$.  Solving for $\alpha$ then becomes:
\begin{equation}
\alpha = \frac{ {\bf t}^T\cdot{\bf M}^{-1}_{SN}\cdot {\bf d} }{ {\bf t}^T\cdot{\bf M}^{-1}_{SN}\cdot {\bf t} } \label{eq:cc_basic}
\end{equation}
To compare multiple template components ${\mathbf t}_j$, e.g. different foregrounds, to a given dataset, the problem becomes a
matrix equation. In the case where we have $N$ different foreground components, we end up with the simple system of linear equations ${\bf Ax}={\bf b}$, where
\begin{equation}
A_{kj}=\mathbf{t}^T_k \cdot \mathbf{M}_{\textrm{SN}}^{-1} \cdot \mathbf{t}_j, \label{ref:eq3}
\end{equation}
\begin{equation}
b_k = \mathbf{t}^T_k \cdot \mathbf{M}_{\textrm{SN}}^{-1} \cdot \mathbf{d}
\end{equation}
\begin{equation}
x_k =\alpha_k.
\end{equation}
When only one template is present, this reduces to equation~\ref{eq:cc_basic} above. 

We assess the uncertainty on our results using the full $N_\mathrm{pix}\times N_\mathrm{pix}$ covariance matrix\footnote{\ch{As $N_\mathrm{pix}$ can be up to $\sim$30,000 for a masked map with $N_\mathrm{side}=64$, this can require up to $\sim14$\,GB of RAM}.}. The CMB anisotropies are taken into account statistically via the covariance matrix, calculated using the theoretical CMB anisotropy power spectrum via $M^S_{ij} = \frac{1}{4\pi} \sum^\infty_{\ell=0} (2\ell+1) C_{\ell} B^2_{\ell} B^2_{\ell, pix} P_{\ell}(\hat{n}_i\cdot\hat{n}_j) $, where $B_{\ell}$ is the beam window function for a Gaussian beam of the appropriate resolution, and $B_{\ell, pix}$ is the HEALPix window function. The power spectrum $C_{\ell}$ is the \WMAP~7-year best-fit theoretical power spectrum \citep{2011Larson}.

We have verified that the uncertainty estimates from the cross-correlation method are robust, both when using the \haslam~and \jonas~templates, by using monte carlo simulations consisting of the foreground maps with realistic template coefficients (i.e. those given in Table \ref{tab:coefficients}); Gaussian noise based on \WMAP~noise levels and $N_\mathrm{obs}$ maps; and CMB realisations generated from the \WMAP~7-year CMB power spectrum. Note that the dominant contribution to the covariance matrix is the CMB.\footnote{\ch{Although these uncertainties could be significantly decreased by subtracting a map of the CMB from the data, this would introduce unquantified uncertainties and aliasing between components depending on the foreground residuals and noise in the CMB map. See Appendix A of \citet{2006Davies} for additional discussion of this issue.}}}

\section{Results} \label{sec:results}
\begin{table*}
\caption{Template coefficients using the \haslam~and \jonas~templates and the KQ75 mask (combined with the 2.3\,GHz mask) for different regions: all-sky (top), North Galactic \ch{hemisphere} (middle) and South Galactic \ch{hemisphere} (bottom). The spectral indices are between the frequency in the row and the one before, except for the synchrotron spectral indices which are between frequency of the dataset and that of the synchrotron template. \ch{We omit the two highest frequency \WMAP~frequency bands as the synchrotron emission at those frequencies is low, and hence noisy. Note that the fractional uncertainty in the synchrotron template coefficients are consistent between the two synchrotron templates, even though the absolute values change.}}
    \tabcolsep 1.5pt
    \small
\begin{tabular}{cccccccccccccc}
\hline
$\nu$ & \multicolumn{6}{c}{\haslam} & & \multicolumn{6}{c}{\jonas}\\
(GHz) & \ha & Dust & Sync & $\beta_\mathrm{sync}$ & $\beta_\mathrm{ff}$ & $\beta_\mathrm{dust}$ &\phantom{    }& \ha & Dust & Sync & $\beta_\mathrm{sync}$ & $\beta_\mathrm{ff}$ & $\beta_\mathrm{dust}$ \\
\hline
22.8 & $10.0\pm0.3$ & $7.9\pm0.2$ & $6.7\pm0.2$ & $-2.96\pm0.01$ & -- &-- & &$10.3\pm0.3$ & $7.4\pm0.2$ & $0.83\pm0.02$ & $-3.11\pm0.01$ & -- &--\\
33.0 & $5.0\pm0.3$ & $2.9\pm0.2$ & $2.0\pm0.2$ & $-2.99\pm0.03$ & $-1.9\pm0.2$ & $-2.7\pm0.2$ & & $5.1\pm0.3$ & $2.7\pm0.2$ & $0.29\pm0.02$ & $-3.07\pm0.02$ & $-1.9\pm0.2$ & $-2.7\pm0.2$\\
40.7 & $3.2\pm0.3$ & $1.7\pm0.2$ & $0.9\pm0.2$ & $-3.02\pm0.05$ & $-2.0\pm0.5$ & $-2.6\pm0.6$ & & $3.3\pm0.3$ & $1.5\pm0.2$ & $0.16\pm0.02$ & $-3.07\pm0.04$ & $-2.0\pm0.5$ & $-2.7\pm0.6$\\
\hline
22.8 & $9.4\pm0.6$ & $7.3\pm0.3$ & $7.2\pm0.4$ & $-2.94\pm0.01$ & -- &-- & &$9.6\pm0.6$ & $6.6\pm0.3$ & $0.93\pm0.03$ & $-3.06\pm0.01$ & -- &--\\
33.0 & $4.5\pm0.6$ & $2.7\pm0.3$ & $2.3\pm0.4$ & $-2.95\pm0.04$ & $-2.0\pm0.4$ & $-2.7\pm0.3$ & & $4.6\pm0.6$ & $2.4\pm0.3$ & $0.35\pm0.03$ & $-3.00\pm0.03$ & $-2.0\pm0.4$ & $-2.7\pm0.4$\\
40.7 & $2.8\pm0.6$ & $1.6\pm0.3$ & $1.2\pm0.4$ & $-2.97\pm0.07$ & $-2.2\pm1.1$ & $-2.4\pm1.1$ & & $2.9\pm0.6$ & $1.4\pm0.3$ & $0.20\pm0.03$ & $-2.98\pm0.04$ & $-2.2\pm1.1$ & $-2.5\pm1.2$\\
\hline
22.8 & $10.2\pm0.4$ & $8.0\pm0.2$ & $6.1\pm0.4$ & $-2.98\pm0.02$ & -- &-- & &$10.1\pm0.4$ & $7.6\pm0.2$ & $0.76\pm0.03$ & $-3.15\pm0.02$ & -- &--\\
33.0 & $5.1\pm0.4$ & $3.0\pm0.2$ & $1.6\pm0.4$ & $-3.04\pm0.06$ & $-1.8\pm0.2$ & $-2.7\pm0.2$ & & $5.1\pm0.4$ & $2.8\pm0.2$ & $0.26\pm0.03$ & $-3.12\pm0.04$ & $-1.8\pm0.2$ & $-2.7\pm0.2$\\
40.7 & $3.4\pm0.4$ & $1.7\pm0.2$ & $0.5\pm0.4$ & $-3.15\pm0.16$ & $-2.0\pm0.6$ & $-2.6\pm0.7$ & & $3.4\pm0.4$ & $1.6\pm0.2$ & $0.13\pm0.03$ & $-3.12\pm0.07$ & $-2.0\pm0.6$ & $-2.7\pm0.7$\\
\hline
\end{tabular}
\label{tab:coefficients}
\end{table*}

Throughout this work, we only consider the area of the sky that has been surveyed at 2.3\,GHz and lies outside of the Galactic plane mask; this is around half of the sky. We discuss the basic results in \ch{Section \ref{sec:basic}}, followed by an investigation of the robustness of these results against various parameters and options in \ch{Section \ref{sec:robustness}}.

Throughout this paper, the units of the template coefficients are $\upmu$K/K for \haslam, $\upmu$K/mK for \jonas, $\upmu$K/R for \ha~and $\upmu$K/$\upmu$K for \fds. All plots use a consistent colour scheme of black lines for \ch{sync}, red lines for \ha~and blue lines for \ch{dust}, with solid lines for \haslam~and dashed lines for \jonas. For display purposes, we have renormalised the \jonas~template coefficients by $(2.326 / 0.408)^{-3.0}/10^{-3}=5.4$, i.e. by assuming a spectral index of $-3.0$ between 408\,MHz and 2.3\,GHz. \ch{We have also offset the \jonas~data points in frequency by 1 per cent.} All error bars are $1 \sigma$.

\subsection{Basic results} \label{sec:basic}
\begin{figure}
\centering
\includegraphics[scale=0.34]{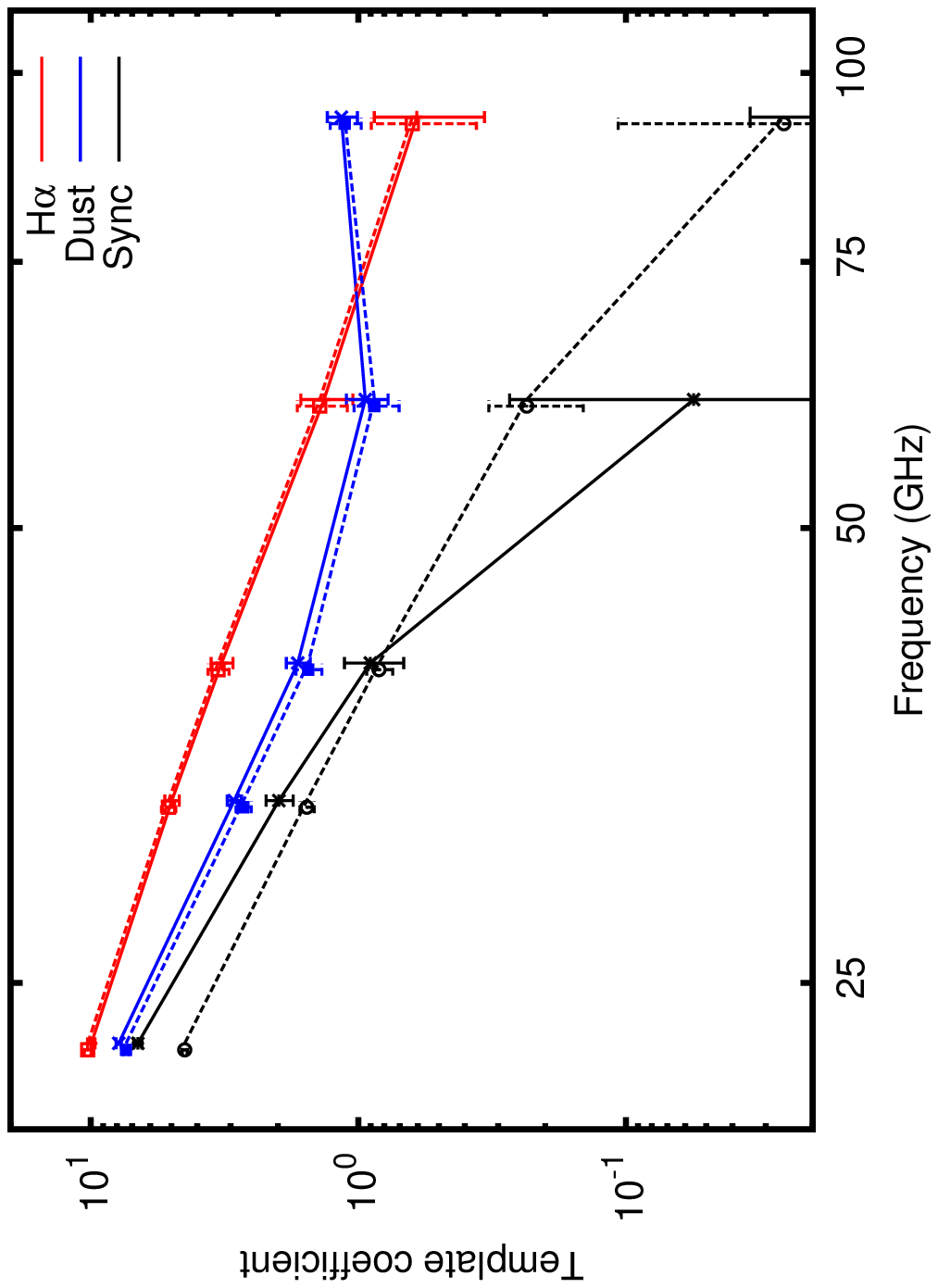}
\caption{Template coefficients for fits to all of the sky observed at 2.3\,GHz outside of the KQ75 mask, using \haslam~(solid lines) and \jonas~(dashed lines). The red data points and lines are for \ha; blue are for dust and black are for synchrotron. There is little difference between the solid and dashed lines for \ddd~and \fds. \ch{Note that only the top of the 93.5\,GHz error bar for the \haslam~analysis is shown here as the central value is negative.}}
\label{fig:results_allsky}
\end{figure}

We apply the cross-correlation method to all 5 individual \WMAP~frequency maps using the \ddd, \fds~and \haslam~templates, and also using the \jonas~template in place of \haslam. The basic results for all of the sky observed at 2.3\,GHz using maps smoothed to 3$^\circ$ and using $N_\mathrm{side}=64$, with the KQ75 mask applied, are shown in Figure \ref{fig:results_allsky}, and are given numerically in Table \ref{tab:coefficients}.

As expected, the synchrotron and free-free template coefficients fall off as a function of frequency (i.e. they have negative spectral indices). The dust-correlated component decreases as a function of frequency where it is tracing the higher frequency side of the AME (see e.g. \citealp{2011PlanckAnom}); it then starts to increase at W-band where the thermal dust component begins to dominate.

There are only a small changes in the \ha~and dust correlation coefficients when the two different synchrotron templates are used, with the \ha~coefficient systematically increasing by around 3 per cent between \haslam~and \jonas, and the dust coefficient systematically decreasing by around 7 per cent. This $\sim3\sigma$ decrease indicates that the synchrotron spectrum is flattening slightly at 2.3\,GHz, however it is not flattening sufficiently to explain all of the dust-correlated emission. For example, based on the rudimentary model described in the introduction, one would expect a $\sim$50 per cent decrease in order for the dust-correlated emission to be due to a flat spectrum synchrotron component with $\beta=-2.5$.

The effect of the flat spectrum component at GHz frequencies can also be seen in the synchrotron spectral indices, where they are steeper between \jonas~and the \WMAP~maps than they are with \haslam; this behaviour requires some flat emission between 408\,MHz and 2.3\,GHz that falls off by \WMAP~frequencies. This is not too surprising, since the average spectral index between \haslam~and \jonas~is \ch{$\beta\sim-2.7$} \citep{2003Platania}\ch{. Given} a spectral index between 408\,MHz and 22.8\,GHz of $-2.96$, a steeper index of $\sim-3.16$ would be expected between 2.3\,GHz and 22.8\,GHz, i.e. similar to what we find here. Note that although calibration errors in the synchrotron maps will affect the spectral indices, any such errors would be expected to be around 10 per cent, and hence would not be an issue here.
% print, alog(exp(-2.96 * alog(0.408/22.8)) / exp(-2.695*alog(0.408/2.3)) ) / alog(2.3/22.8) -->     -3.15979

That the differences in the template coefficients when using the \jonas~or \haslam~templates are small confirms that the bulk of the diffuse anomalous emission seen by \WMAP~is indeed dust-correlated, and hence is more likely to be spinning dust rather than synchrotron emission with a flattening spectral index. The dust-correlated emission at 22.8\,GHz is detected at a significance level of $>30\sigma$.

\subsection{Robustness} \label{sec:robustness}
\begin{figure*}
\centering
\includegraphics[scale=0.34]{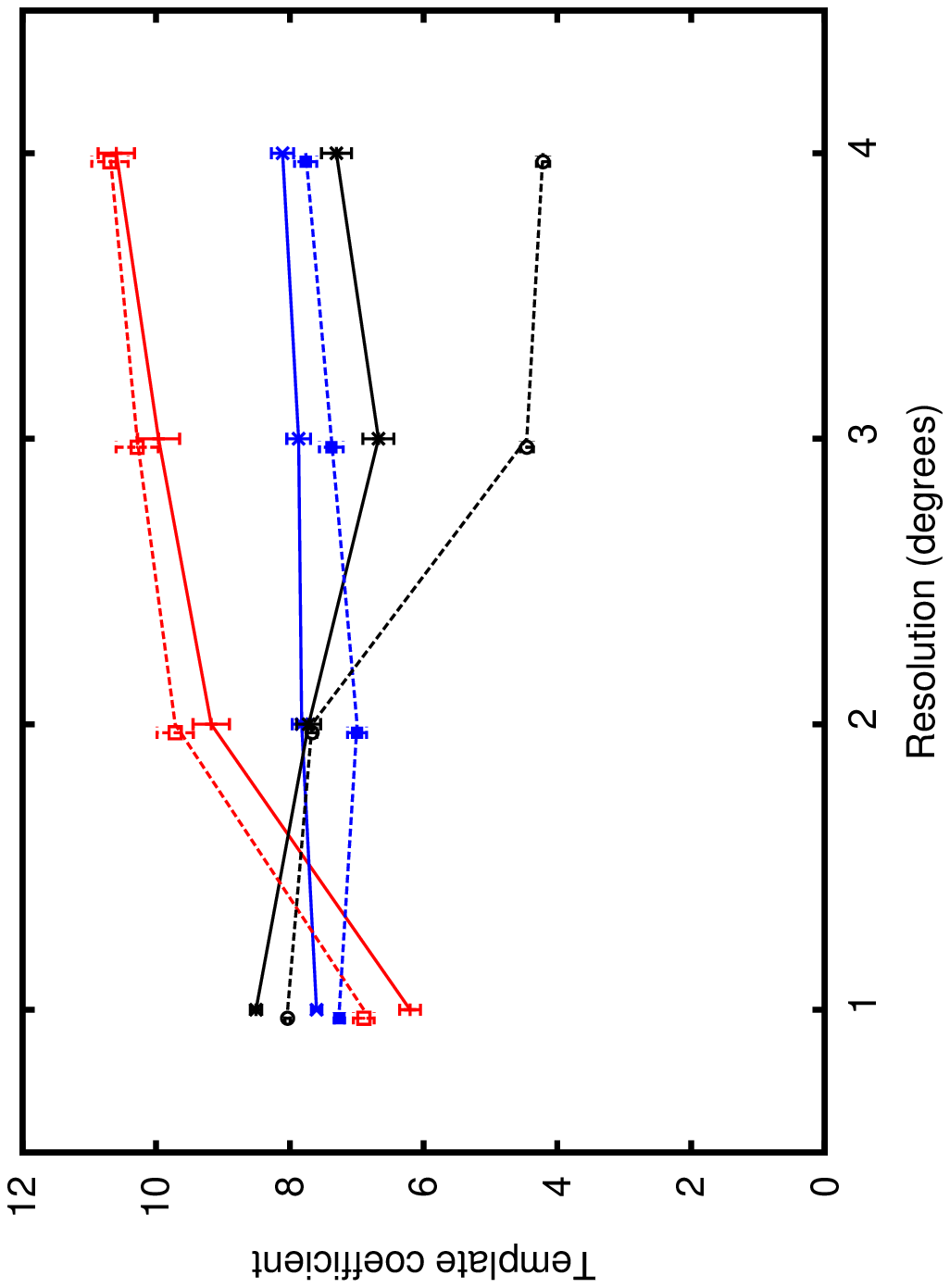}
\includegraphics[scale=0.34]{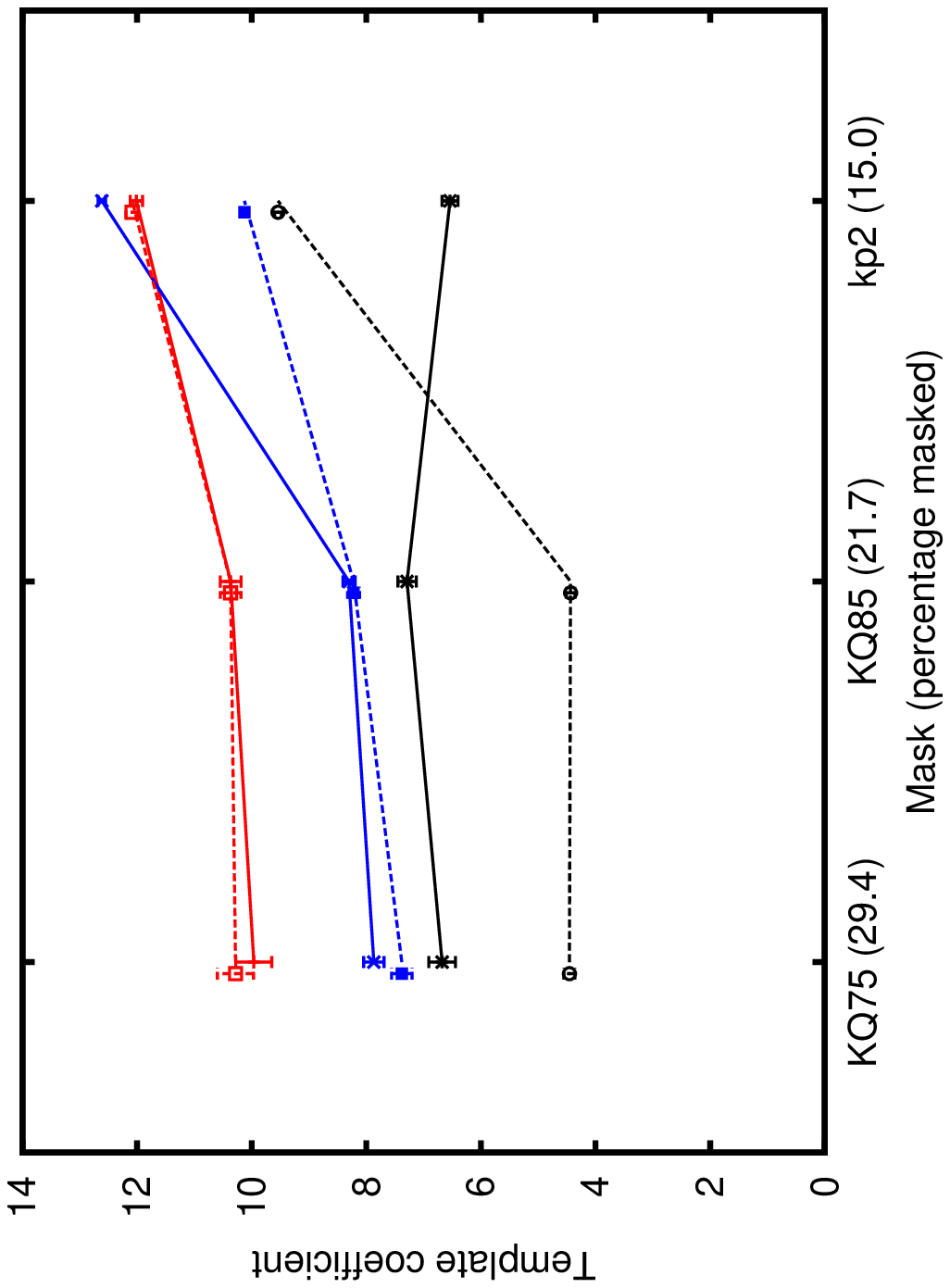}
\includegraphics[scale=0.34]{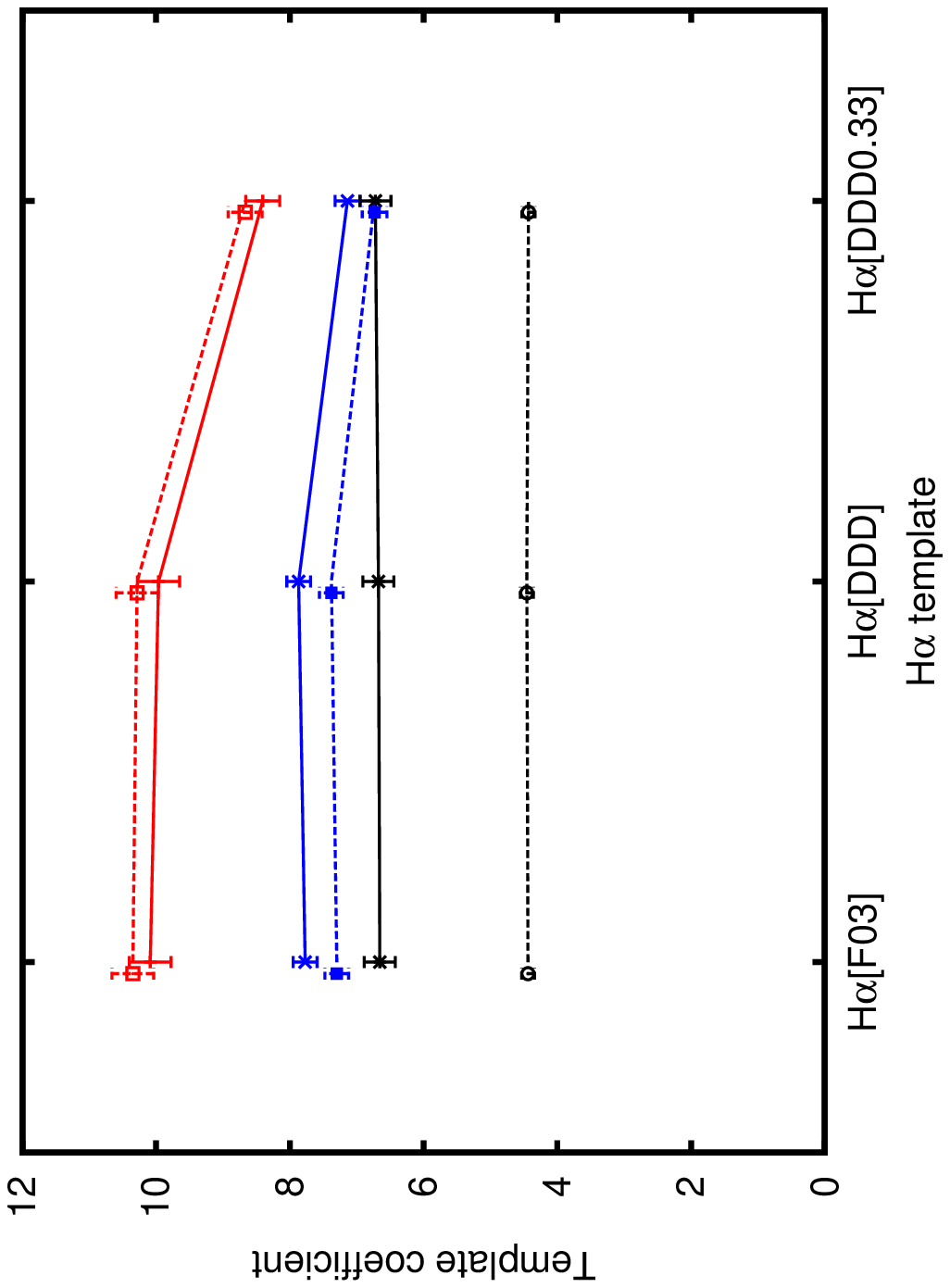}
\includegraphics[scale=0.34]{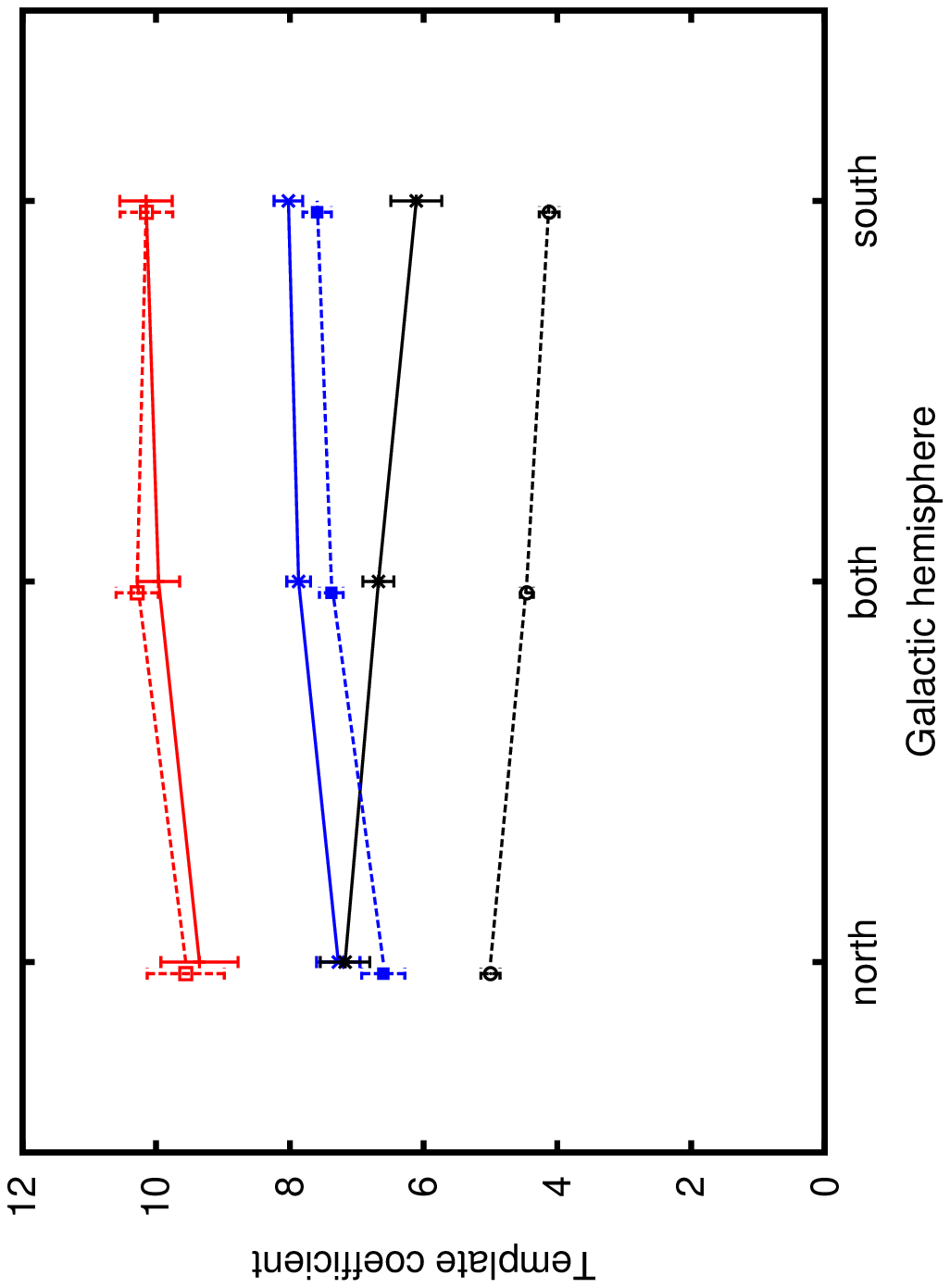}
\caption{22.8\,GHz Template coefficients vs. resolution for all sky, showing convergence by 3$^\circ$ (top-left); vs. the \WMAP~galactic cut mask used (top-right); vs. the \ha~template for all sky (bottom-left); and for all sky compared with North and South Galactic \ch{hemispheres} (bottom-right). All panels use black lines for synchrotron, red lines for \ha~and blue lines for dust, with solid lines indicating results using \haslam, and dashed lines those using \jonas. \ch{The lines between data points are included to guide the eye as to the trends of the data rather than being physically meaningful.}}
\label{fig:robustness}
\end{figure*}
To determine the robustness of our results, we look at the effects of different resolutions, foreground masks \ch{(including looking at the North and South Galactic \ch{hemispheres} separately), \ha~templates and} using both synchrotron templates together. The results from these tests are described \ch{below, and are plotted in Figure \ref{fig:robustness} with the template coefficients presented in Table \ref{tab:robustnesscoefficients}. We} find that our results as described above are robust.

\begin{table}
\caption{Table of synchrotron spectral indices using the KQ75 mask and resolutions of 1$^\circ$, 2$^\circ$, 3$^\circ$ and 4$^\circ$. Top: using \haslam. Bottom: using \jonas.}
    \tabcolsep 3.5pt
    \small
\begin{tabular}{ccccc}
\hline
$\nu$ \ch{(GHz)} & 1$^\circ$ & 2$^\circ$ & 3$^\circ$ & 4$^\circ$\\
\hline
22.8 & $-2.90\pm0.01$ & $-2.93\pm0.01$ & $-2.96\pm0.01$ & $-2.94\pm0.01$\\
33.0 &  $-2.90\pm0.01$ & $-2.92\pm0.02$ & $-2.99\pm0.03$ & $-2.97\pm0.02$\\
40.7 & $-2.91\pm0.01$ & $-2.93\pm0.03$ & $-3.02\pm0.05$ & $-3.00\pm0.05$\\
\hline
22.8 & $-2.85\pm0.01$ & $-2.87\pm0.01$ & $-3.11\pm0.01$ & $-3.13\pm0.01$\\
33.0 & $-2.81\pm0.01$ & $-2.84\pm0.01$ & $-3.07\pm0.02$ & $-3.10\pm0.02$\\
40.7 & $-2.79\pm0.01$ & $-2.82\pm0.01$ & $-3.07\pm0.04$ & $-3.08\pm0.04$\\
\hline
\end{tabular}
\label{tab:syncspectrum}
\end{table}

\begin{itemize}
\item {\bf Resolution:} We have repeated the analysis above using maps smoothed to \ch{1$^\circ$, 2$^\circ$ and 4$^\circ$} to look at the effects on the template coefficients due to different map resolutions; as shown by \citet{2011Ghosh} these can be considerable. The top-left panel of Figure \ref{fig:robustness} shows the template coefficients at 22.8\,GHz vs. the resolution of the map. Importantly, the dust coefficient remains relatively stable regardless of resolution. The \ha~template coefficient increases with larger beam size, with the  H$\alpha$ coefficients at 3$^\circ$ being consistent with a free-free electron temperature of $6000-7000$\,K, unlike previous studies such as \citet{2006Davies}. This is likely to be due to a number of effects including under-sampling reducing the smaller-scale power, stellar residuals, baseline offsets, etc.; see \citet{2011Ghosh} for discussion.

The synchrotron coefficient decreases with larger beam size, potentially due to contaminating compact regions (which typically have flatter spectral indices) whose effect at larger resolutions has been reduced by the smoothing. The effect that this has on the synchrotron spectral index is shown in Table \ref{tab:syncspectrum} -- this can be quite large, particularly for 2.3\,GHz\ch{. There} is a trend towards steeper spectral indices at coarser resolutions. These trends are present when using either \haslam~or \jonas. The coefficients have however, converged by $\sim3^\circ$, verifying our choice of this resolution in the above analysis.

\item {\bf Mask:} In order to check that our results are robust against the different masks that could be applied, and hence the different amount of strong Galactic plane emission that is masked out, we compare our results from the KQ75 mask at 22.8\,GHz with those from the KQ85 and kp2 masks (as described in section \ref{sec:wmap7}). The results are shown in the top-right panel in Figure \ref{fig:robustness}. The template coefficients increase as the size of the mask is decreased thereby including more of the Galactic plane. This is particularly the case for the \ha~map, as would be expected if emission is being absorbed along the Galactic plane. There is a marked difference in the dust template for kp2 between \haslam~and \jonas, and a corresponding increase in the \jonas~coefficient, which is likely due to the presence of flatter synchrotron spectra closer in to the Galactic plane. However, off the Galactic plane the results remain robust between \haslam~and \jonas.

\ch{We also} consider the north- and south-Galactic hemispheres separately as well as combined; the results are shown in the bottom-right panel of Figure \ref{fig:robustness}, and the numerical results are also given in Table \ref{tab:coefficients}. Although the results are largely consistent with each other, we find a slight asymmetry: the south Galactic \ch{hemispheres} have a lower synchrotron correlation, and higher dust and \ha~correlations, than the northern hemisphere. Although this is different from the results in \citet{2011Ghosh}, this is likely due to the exclusion of much of the Northern hemisphere from the analysis due to it not being covered by the 2.3\,GHz map.

\item {\bf \ha~template:} We use the three different \ha~templates to look for the dependency of our results on the accuracy of these maps. The results from analyses using both the \fink map and \ddd~maps with and without dust corrections are shown in the bottom-left panel of Figure \ref{fig:robustness}. We find that the dust correction significantly decreases the coefficient of free-free emission, and to a lesser extent the thermal dust emission, but that this equally affects the analyses using \haslam~and \jonas.

\item {\bf Multiple synchrotron templates:} We also use all four templates simultaneously to fit the \WMAP~data. Although the resulting synchrotron template coefficients are meaningless due to the large amount of degeneracy between the spatial structure in the two synchrotron templates, it is a useful cross-check for the \ha~and dust coefficients. We find that this does not significantly affect our \ch{results. For example, at 22.8\,GHz we find the template coefficient for H$\alpha$ when using just \haslam~to be $10.0 \pm 0.3$\,$\upmu$K/R and that for dust to be $7.9\pm0.2$\,$\upmu$K/$\upmu$K. The coefficients change to $10.3\pm0.3$\,$\upmu$K/R and $7.3\pm0.2$\,$\upmu$K/$\upmu$K respectively when including both \haslam~and \jonas, and to $10.3\pm0.3$\,$\upmu$K/R and $7.4\pm0.2$\,$\upmu$K/$\upmu$K when just using \jonas.} In all cases, the dust coefficient varies by less than 10 per cent.

\end{itemize}

\subsection{Results in specific regions} \label{sec:regions}
\begin{figure}
\centering
\includegraphics[scale=0.3]{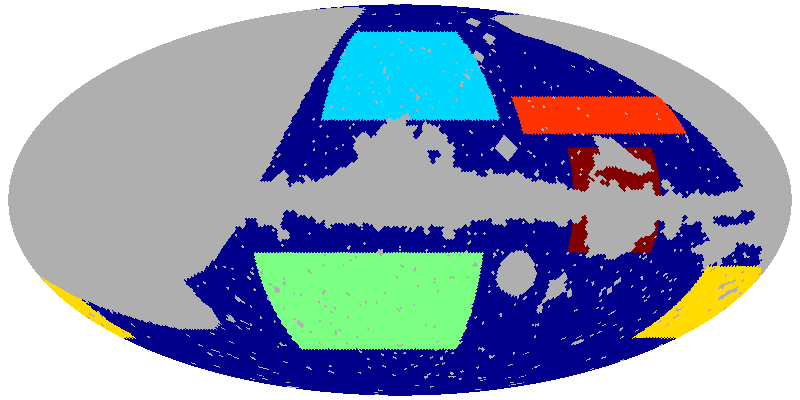}
\caption{Full-sky map showing the individual regions considered at here, along with the 2.3\,GHz survey area mask and the \WMAP~7-year temperature mask (light dark). The regions are North Polar Spur (top; blue), Gum nebula (middle-right; red), Eridanus region (bottom-right and -left; yellow), South Galactic latitudes (bottom-middle; green) and the Northern region (top-right; orange)}
\label{fig:masks}
\end{figure}

\begin{figure*}
\centering
\includegraphics[scale=0.34]{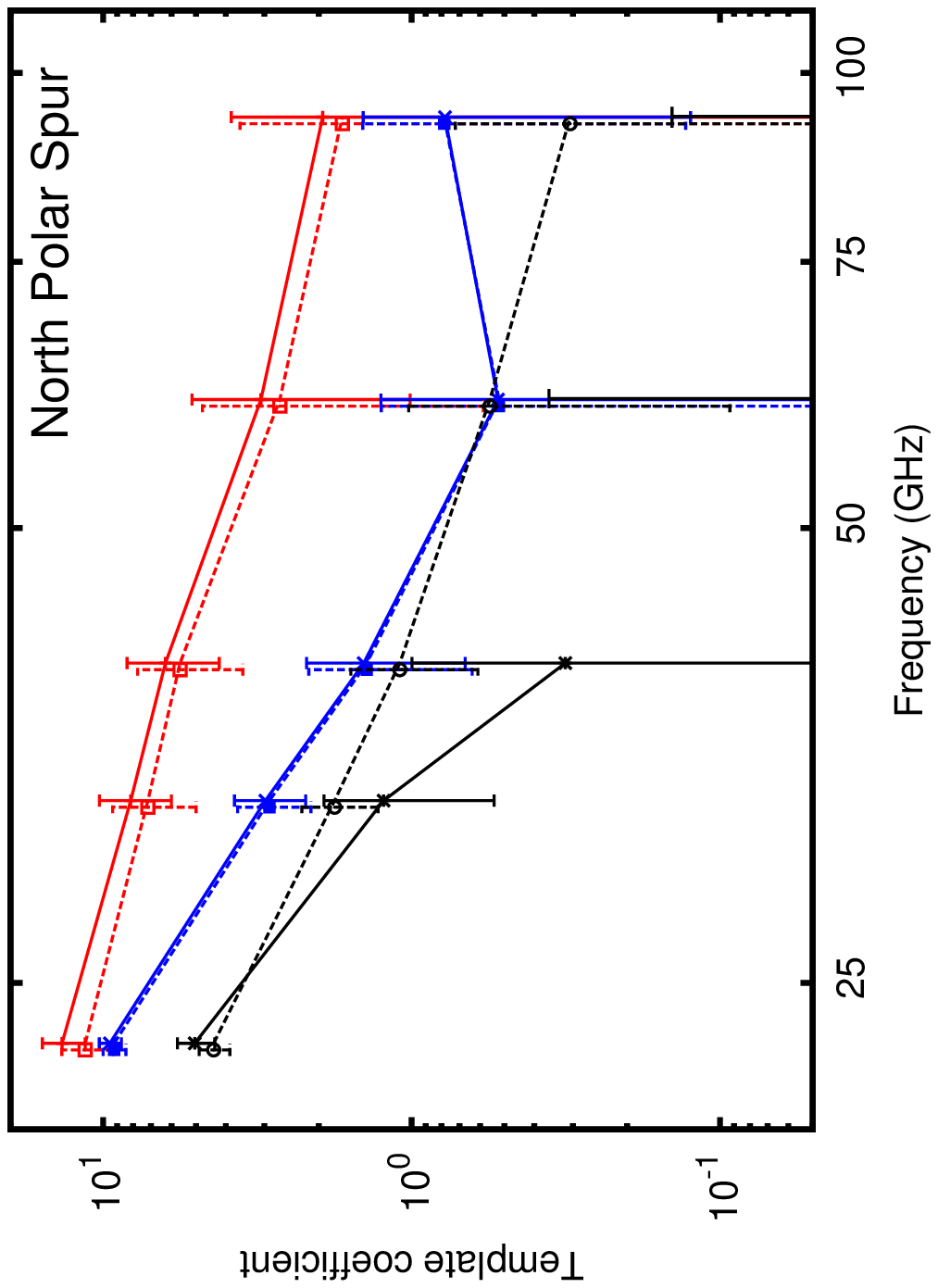}
\includegraphics[scale=0.34]{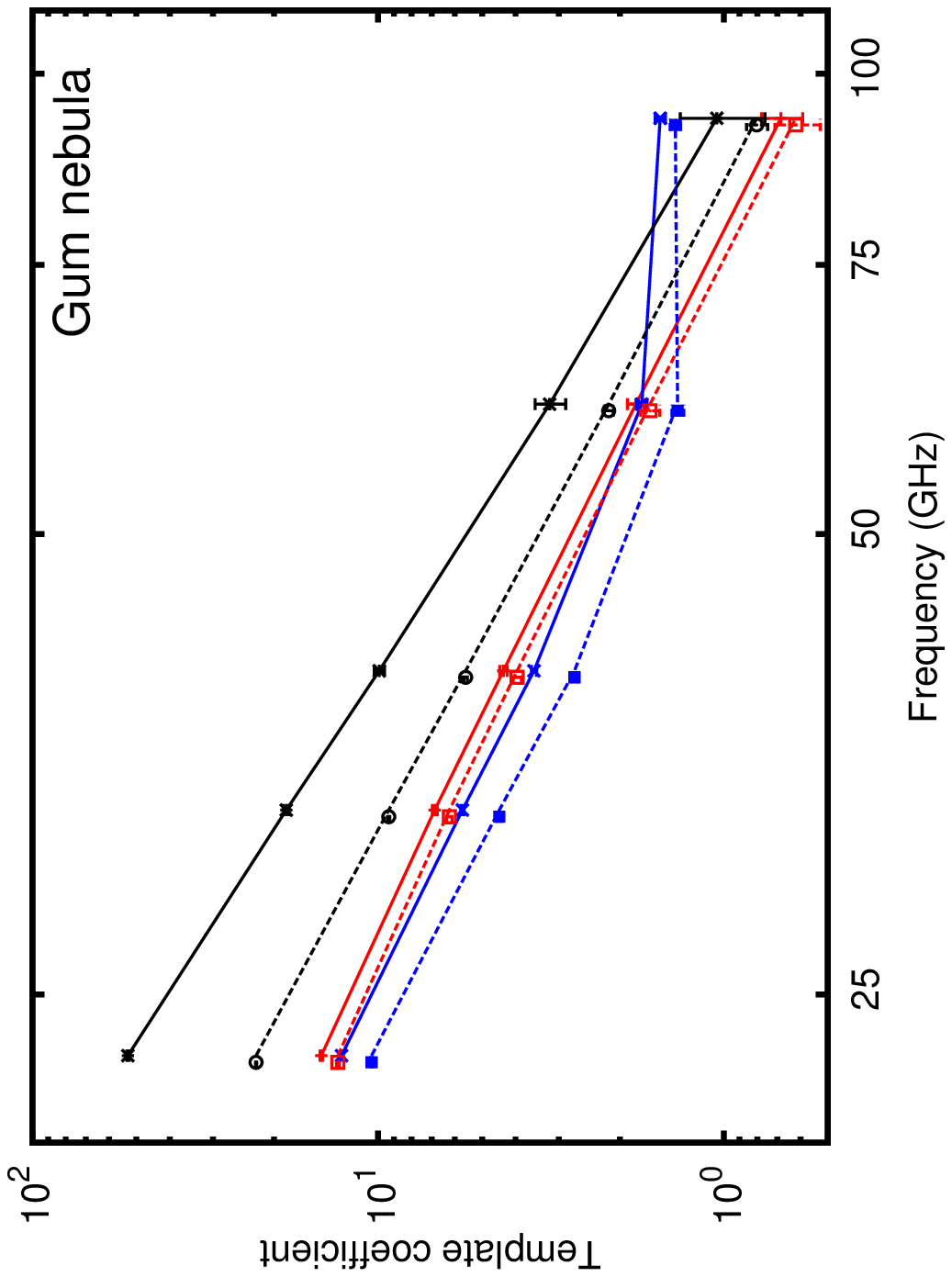}
\includegraphics[scale=0.34]{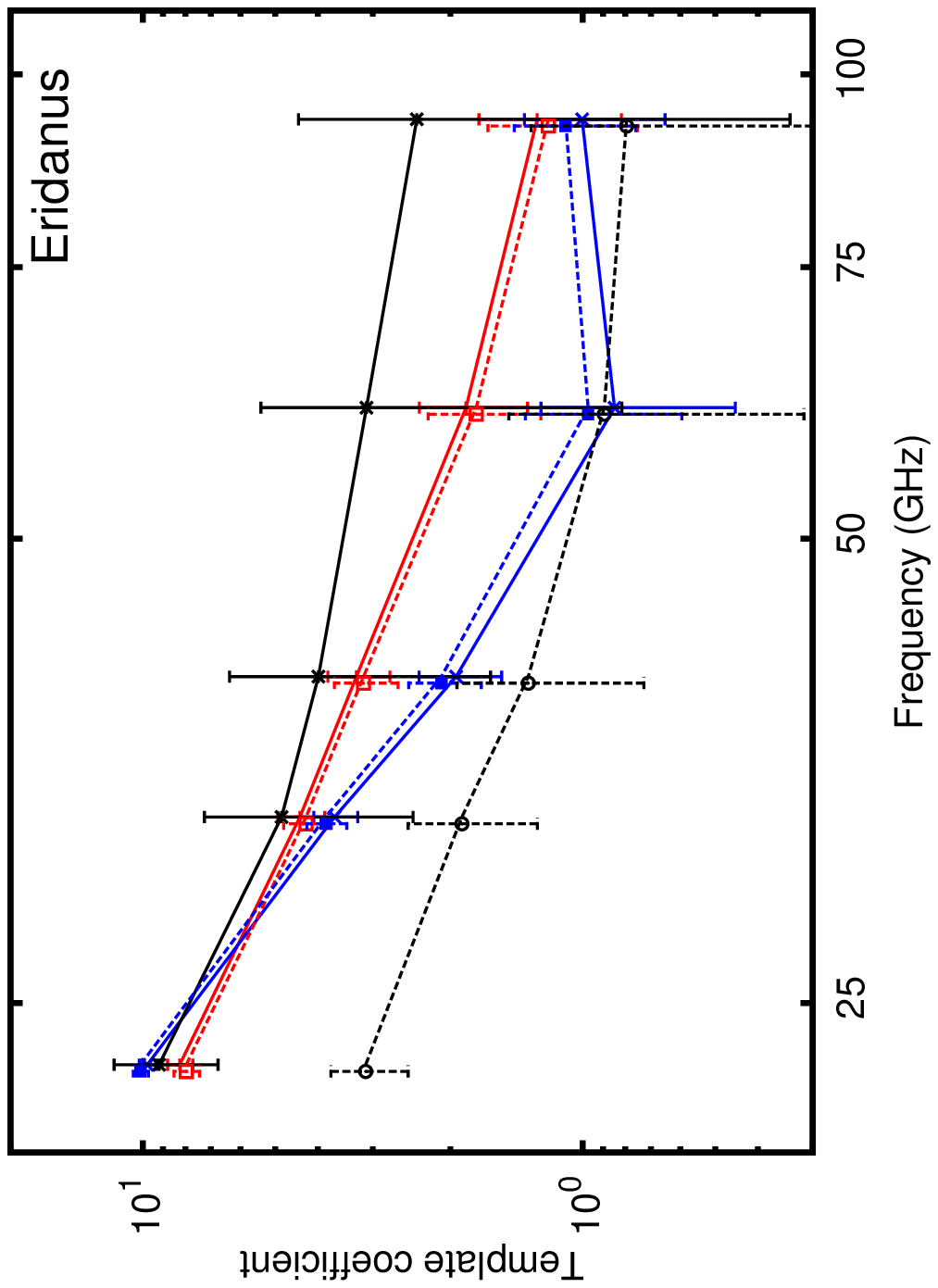}
\includegraphics[scale=0.34]{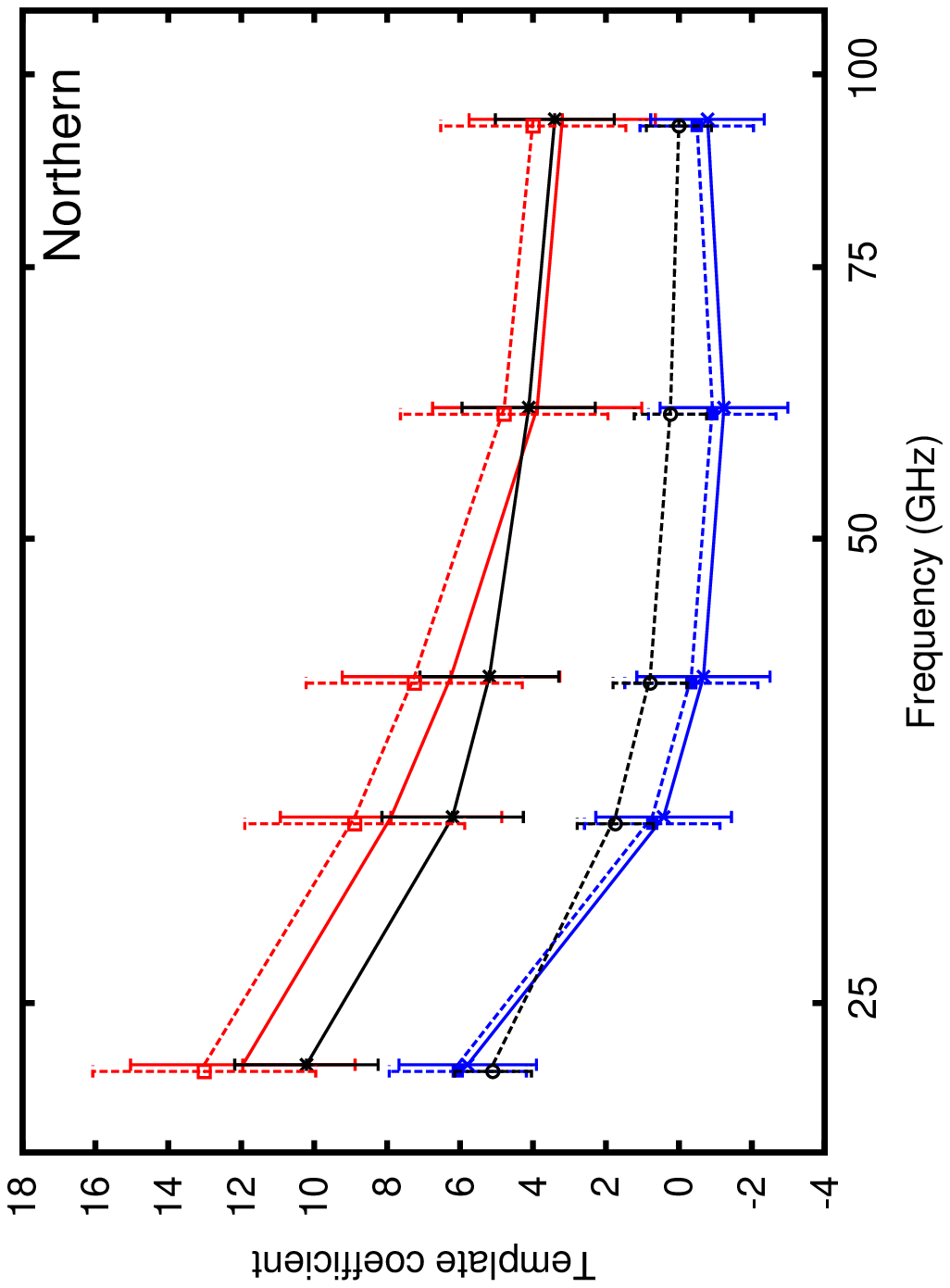}
\includegraphics[scale=0.34]{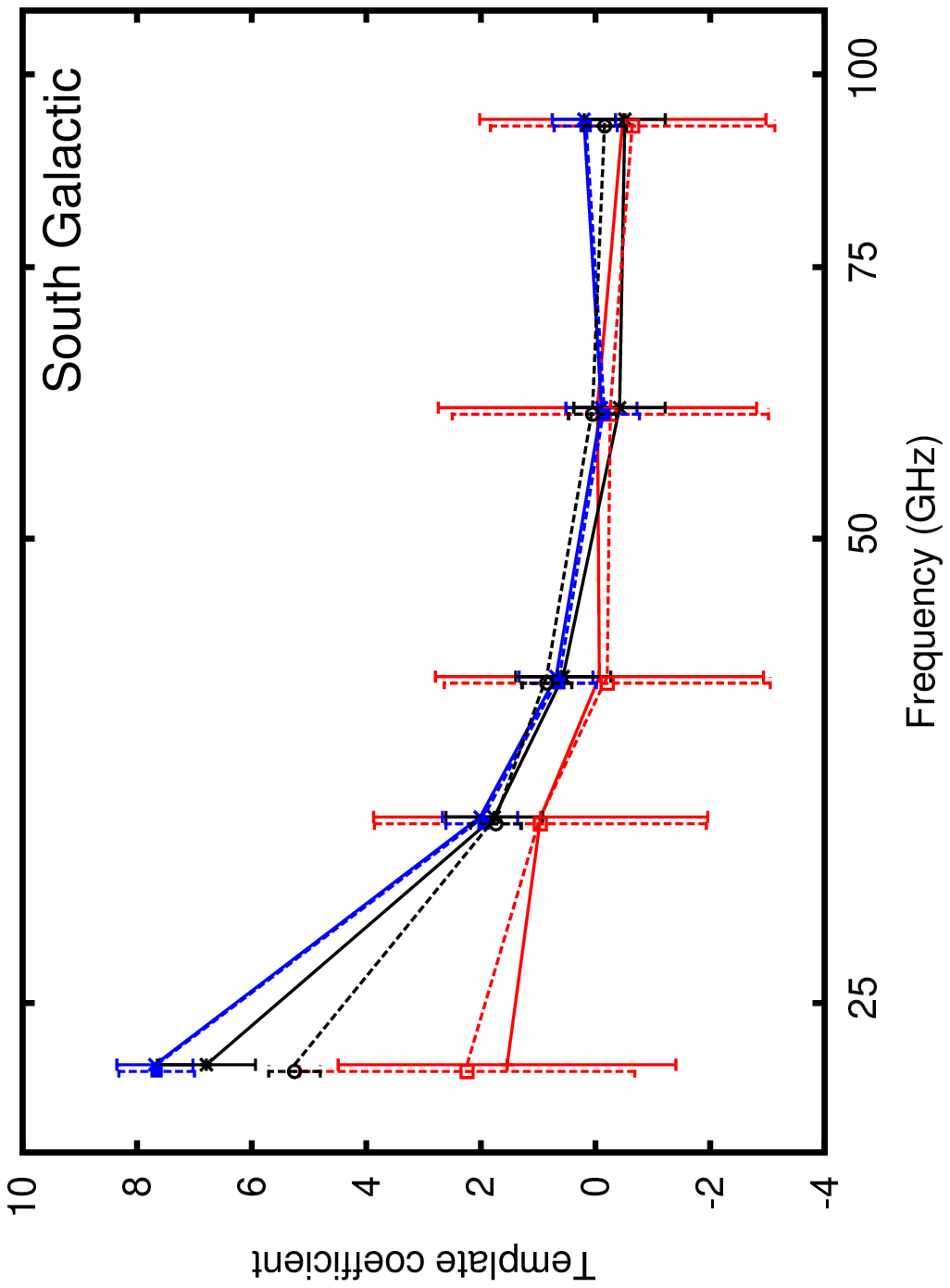}
\caption{Template coefficients for fits using Haslam 408\,MHz (solid lines) and Jonas 2.3\,GHz (dashed lines). The red data points and lines are for \ha; blue are for dust and black are for synchrotron. Top-left is for the north polar spur; top-right for the Gum nebula; middle-left for Eridanus; middle-right for the Northern region; bottom for the South Galactic regions.}
\label{fig:region_coefficients}
\end{figure*}

\ch{Although our results are robust on average across the entire sky, it is also important to look at their robustness in different Galactic environments. This also lets us test the effect of variations in the synchrotron spectral index across the sky. To do this, we focus on a set of five distinct regions, as outlined in Figure \ref{fig:masks}, in a similar way to \citet{2006Davies}. Since we are using a larger beam size than in \citet{2006Davies}, we use significantly larger regions.  This gives us more independent measures of the CMB fluctuations as well as allowing us to trace enough of the morphology to distinguish between the different components. \citet{2011Ghosh} look at a larger number of regions, based on the template morphology, which overlap with, but are distinct from, the regions we look at here.}

The results from applying template fitting to these five regions are shown in Figure \ref{fig:region_coefficients}\ch{, with the template coefficients presented in Table \ref{tab:regioncoefficients}.} Although the uncertainties for each region are a lot higher, due to the smaller area for each region, the dust and free-free coefficients remain consistent within the uncertainties in all cases. The results from the individual regions are described below.

\begin{itemize}
\item {\bf North Polar Spur} ($310^\circ \geq l \leq 40^\circ$, $30^\circ \leq b \leq 70^\circ$). This region is dominated by an arc of steep spectrum synchrotron radiation that is thought to be due to the remnant of a nearby supernova (see e.g. \citealp{2007Wolleben}). The region also contains a spur of thermal dust emission (see Figure \ref{fig:templates}).

The results are shown in the top-left panel of Figure \ref{fig:region_coefficients}. Although there are large uncertainties on the template coefficients, there is very little difference in the \ha~and dust template coefficients between \haslam~and \jonas.

\item {\bf Gum nebula} ($240^\circ\leq l \leq280^\circ$, $-20^\circ \leq b \leq 20^\circ$). This region is dominated by free-free emission, with a reduced amount of synchrotron \ch{emission. It} also contains a significant amount of thermal dust emission (see Figure \ref{fig:templates}). We mask out the central part of the nebula, and look at the surrounding emission only, to avoid issues with the \ha~template in areas of high dust absorption.

The results are shown in the top-right panel of Figure \ref{fig:region_coefficients}. The uncertainties are smaller since this is a bright region. We find a significantly increased synchrotron template coefficient compared to the rest of the sky -- approximately 4 times higher -- however this will be dominated by the uncertainty of the free-free correction to these templates. The \ha~and dust coefficients between the two synchrotron templates differ, with both being reduced when using \jonas. This implies that there is a reasonable amount of flat spectrum synchrotron emission in this region, but the result will depend on the free-free corrections to the synchrotron templates.

\item {\bf Eridanus} ($170^\circ\leq l \leq210^\circ$, $-55^\circ\leq b \leq-25^\circ$). The main feature in this region is a line of thermal dust and free-free emission (as can be seen in Figure \ref{fig:templates}), with a reduced amount of synchrotron emission compared to most of the sky.

The template coefficients are shown in the middle-left panel of Figure \ref{fig:region_coefficients}. The dust and \ha~template coefficients from analyses using \haslam~and \jonas~are consistent with each other, however the uncertainties in this region are large as the foreground emission is small (in particular, the synchrotron emission is low) and hence the region is dominated more by CMB emission.

\item {\bf Northern region} ($220^\circ\leq l \leq300^\circ$, $25^\circ\leq b \leq40^\circ$). This is a region of low foreground emission in the northern Galactic hemisphere, just above the Gum nebula.

The results are shown in the middle-right panel of Figure \ref{fig:region_coefficients} are consistent between \haslam~and \jonas (with a large uncertainty due to the low foregrounds), with a slightly higher dust coefficient when \jonas~is used rather than \haslam.

\item {\bf South Galactic latitudes} ($320^\circ\geq l \leq70^\circ$, $-60^\circ\leq b \leq-20^\circ$). This is a region to the south of the Galactic centre that contains moderate levels of foreground emission, primarily synchrotron and thermal dust emission, with a lower level of free-free emission. Figure \ref{fig:templates} shows that this region has not been well-mapped by \ha~surveys, with clear mottling present due to the field-of-view of the observations.

The results are shown in the bottom panel of Figure \ref{fig:region_coefficients}. We find coefficients for the \ha~template that are consistent with zero within \ch{large} uncertainties, confirming the presence of issues with the \ha~template in this region. There are high dust and synchrotron coefficients, which are consistent with each other within the uncertainties.

\end{itemize}

\subsection{\ch{Template cross-correlations}}
\begin{table}
\caption{\ch{Template cross-correlation matrices for the KQ75 mask (combined with the 2.3GHz mask), for all of the sky (top), the Northern region (middle) and the Eridanus region (bottom).}}
    \tabcolsep 1.5pt
    \small
\begin{tabular}{cccccc}
\hline
 & \ddd & \fds & \haslam & \jonas & const.\\
\hline
\ddd & 1.00 & 0.66 & 0.64 & 0.59 & 0.67\\
\fds & 0.66 & 1.00 & 0.79 & 0.75 & 0.79\\
\haslam & 0.64 & 0.79 & 1.00 & 0.93 & 0.96\\
\jonas & 0.59 & 0.75 & 0.93 & 1.00 & 0.89\\
const. & 0.67 & 0.79 & 0.96 & 0.89 & 1.00\\
\hline
\ddd & 1.00 & 0.40 & 0.45 & 0.39 & 0.42\\
\fds & 0.40 & 1.00 & 0.61 & 0.54 & 0.63\\
\haslam & 0.45 & 0.61 & 1.00 & 0.86 & 0.91\\
\jonas & 0.39 & 0.54 & 0.86 & 1.00 & 0.80\\
const. & 0.42 & 0.63 & 0.91 & 0.80 & 1.00\\
\hline
\ddd & 1.00 & 0.33 & 0.26 & 0.32 & 0.28\\
\fds & 0.33 & 1.00 & 0.44 & 0.47 & 0.52\\
\haslam & 0.26 & 0.44 & 1.00 & 0.95 & 0.88\\
\jonas & 0.32 & 0.47 & 0.95 & 1.00 & 0.87\\
const. & 0.28 & 0.52 & 0.88 & 0.87 & 1.00\\
\hline
\end{tabular}
\label{tab:correlations}
\end{table}

\ch{Correlations between templates can result in degeneracies between the different template coefficients, which are important to consider in this type of analysis. The correlations can be calculated from the off-diagonal terms of ${\mathbf A}$ from Equation \ref{ref:eq3} via
\begin{equation}
C_{ij} = A_{ij} / \sqrt{A_{ii} A_{jj}}.
\end{equation}
Table \ref{tab:correlations} gives the correlation matrices between the different types of templates for the whole of the area of sky used here, as well as in two illustrative regions (Northern and Eridanus).

As expected, in all cases the two synchrotron templates are highly correlated with each other, but not completely. Typical values for the correlation between the synchrotron templates are around 0.9. On the whole sky, there are significant correlations between all of the templates, due to the similar large-scale morphology of the Galactic plane and lower latitude regions. There is also a high degree of correlation with the constant template. The correlations decrease when looking at smaller regions of sky, where the large-scale structure of our Galaxy is less significant. That the correlation coefficients vary between different regions, whilst the regions have the same results, indicates that these correlations do not present a significant issue. This issue has been explored in more detail in Appendix B of \citet{2006Davies}.

An improved approach to take here would be to filter the maps and the templates to remove large-scale modes, and to concentrate on the smaller scale fluctuations, which will be investigated in a future work.}

\section{Conclusions} \label{sec:conclusion}

We have applied the template fitting method to the \WMAP~7-year maps, using synchrotron templates at 408\,MHz and 2.3\,GHz. We find that using the higher frequency synchrotron template slightly reduces the dust-correlated component across the whole sky by up to $10$ per cent, which indicates that the bulk of the emission is much more likely to be due to spinning dust emission rather than flattening synchrotron radiation, in agreement with the results of e.g. \citet{2004deOliveiraCosta,2011PlanckAnom}. We find that our results are robust against different choices of the map resolution, Galactic masks and \ha~template, and that our uncertainty estimates are robust. We find a slight asymmetry in the template coefficients between the north and south Galactic \ch{hemispheres}, and also different template coefficients in different regions of the \ch{sky. However} we find that the dust coefficient remains robust when using either synchrotron template in all cases off the Galactic plane.

In the future, data from \Planck~LFI will provide additional measurements that can be compared with the templates to help pin down the spectra of the different components. A major step forward will be when the 5\,GHz all-sky map currently being produced by C-BASS becomes available; as this is all-sky, and at twice the frequency of \jonas, this will greatly improve the constraints on flat synchrotron emission at higher frequencies than those considered here, and will further help pin down the cause of the anomalous microwave emission.

\section*{Acknowledgments}
We acknowledge the use of the Legacy Archive for Microwave Background Data Analysis (LAMBDA). Support for LAMBDA is provided by the NASA Office of Space Science. \ch{Some of the results in this paper have been derived using the HEALPix \citep{2005Gorski} package.} CD acknowledges an STFC Advanced Fellowship and an ERC grant under the FP7. \ch{This research was supported by the Agence Nationale de Recherche (ANR-08-CEXC-0002-01).} \ch{We thank the anonymous reviewer for their rigorous, constructive comments.}

\appendix
\section{Template coefficients}
\begin{table*}
\caption{\ch{Template coefficients at 22.8\,GHz as per Figure \ref{fig:robustness}, including the corresponding spectral index for synchrotron.}}
    \tabcolsep 2.5pt
    \small
\begin{tabular}{cccccccccc}
\hline
 & \multicolumn{4}{c}{\haslam} & & \multicolumn{4}{c}{\jonas}\\
Identifier & \ha & Dust & Sync & $\beta_\mathrm{sync}$ & \phantom{    }& \ha & Dust & Sync & $\beta_\mathrm{sync}$ \\
\hline
1$^\circ$ & $6.2\pm0.1$ & $7.6\pm0.2$ & $8.5\pm0.1$ & $-2.90\pm0.01$ & & $6.9\pm0.1$ & $7.3\pm0.2$ & $1.5\pm0.1$ & $-2.85\pm0.01$\\
2$^\circ$ & $9.2\pm0.2$ & $7.8\pm0.3$ & $7.7\pm0.1$ & $-2.93\pm0.01$ & & $9.7\pm0.3$ & $7.0\pm0.3$ & $1.4\pm0.1$ & $-2.87\pm0.01$\\
3$^\circ$ & $10.0\pm0.6$ & $7.9\pm0.3$ & $6.7\pm0.2$ & $-2.96\pm0.01$ & & $10.3\pm0.7$ & $7.4\pm0.3$ & $0.8\pm0.2$ & $-3.11\pm0.01$\\
4$^\circ$ & $10.6\pm0.8$ & $8.1\pm0.3$ & $7.3\pm0.2$ & $-2.94\pm0.01$ & & $10.7\pm0.8$ & $7.8\pm0.3$ & $0.8\pm0.2$ & $-3.13\pm0.01$\\
\hline
KQ75 & $10.0\pm0.6$ & $7.9\pm0.3$ & $6.7\pm0.2$ & $-2.96\pm0.01$ & & $10.3\pm0.7$ & $7.4\pm0.3$ & $0.8\pm0.2$ & $-3.11\pm0.01$ \\
KQ85 & $10.4\pm0.9$ & $8.3\pm0.2$ & $7.3\pm0.1$ & $-2.94\pm0.01$ & & $10.4\pm0.9$ & $8.2\pm0.2$ & $0.8\pm0.1$ & $-3.11\pm0.01$ \\
kp2 & $12.0\pm0.8$ & $12.6\pm0.1$ & $6.5\pm0.1$ & $-2.97\pm0.01$ & & $12.1\pm0.8$ & $10.1\pm0.1$ & $1.8\pm0.1$ & $-2.78\pm0.01$ \\
\hline
\fink & $10.1\pm0.6$ & $7.8\pm0.3$ & $6.7\pm0.2$ & $-2.96\pm0.01$ & & $10.3\pm0.7$ & $7.3\pm0.3$ & $0.8\pm0.2$ & $-3.11\pm0.01$ \\
\ddd & $10.0\pm0.6$ & $7.9\pm0.3$ & $6.7\pm0.2$ & $-2.96\pm0.01$ & & $10.3\pm0.7$ & $7.4\pm0.3$ & $0.8\pm0.2$ & $-3.11\pm0.01$ \\
\dddfd & $8.4\pm0.6$ & $7.1\pm0.3$ & $6.7\pm0.2$ & $-2.96\pm0.01$ & & $8.7\pm0.7$ & $6.7\pm0.3$ & $0.8\pm0.2$ & $-3.11\pm0.01$ \\
\hline
north & $9.4\pm0.4$ & $7.3\pm0.6$ & $7.2\pm0.3$ & $-2.94\pm0.01$ & & $9.6\pm0.5$ & $6.6\pm0.6$ & $0.9\pm0.3$ & $-3.06\pm0.01$ \\
both & $10.0\pm0.6$ & $7.9\pm0.3$ & $6.7\pm0.2$ & $-2.96\pm0.01$ & & $10.3\pm0.7$ & $7.4\pm0.3$ & $0.8\pm0.2$ & $-3.11\pm0.01$ \\
south & $10.2\pm0.5$ & $8.0\pm0.4$ & $6.1\pm0.2$ & $-2.98\pm0.02$ & & $10.1\pm0.6$ & $7.6\pm0.4$ & $0.8\pm0.2$ & $-3.14\pm0.02$ \\
\hline
\end{tabular}
\label{tab:robustnesscoefficients}
\end{table*}

\begin{table*}
\caption{\ch{Template coefficients, as per Table \ref{tab:coefficients}, for the five regions shown in Figure \ref{fig:region_coefficients}. From top to bottom: North Polar Spur, Gum nebula, Eridanus, Northern and South Galactic.}}
    \tabcolsep 1.5pt
    \small
\begin{tabular}{cccccccccccccc}
\hline
$\nu$ & \multicolumn{6}{c}{\haslam} & & \multicolumn{6}{c}{\jonas}\\
(GHz) & \ha & Dust & Sync & $\beta_\mathrm{sync}$ & $\beta_\mathrm{ff}$ & $\beta_\mathrm{dust}$ &\phantom{    }& \ha & Dust & Sync & $\beta_\mathrm{sync}$ & $\beta_\mathrm{ff}$ & $\beta_\mathrm{dust}$ \\
\hline
22.8 & $13.6\pm2.2$ & $9.5\pm0.8$ & $5.0\pm0.7$ & $-3.03\pm0.03$ & -- &-- & &$11.4\pm2.2$ & $9.2\pm0.8$ & $0.81\pm0.09$ & $-3.12\pm0.05$ & -- &-- \\
33.0 & $8.1\pm2.1$ & $3.0\pm0.8$ & $1.2\pm0.7$ & $-3.10\pm0.13$ & $-1.4\pm0.8$ & $-3.1\pm0.7$ & & $7.2\pm2.2$ & $2.9\pm0.8$ & $0.33\pm0.09$ & $-3.02\pm0.10$ & $-1.3\pm1.0$ & $-3.1\pm0.8$\\
40.7 & $6.3\pm2.1$ & $1.4\pm0.8$ & $0.3\pm0.7$ & $-3.25\pm0.47$ & $-1.2\pm2.0$ & $-3.5\pm2.8$ & & $5.6\pm2.1$ & $1.4\pm0.8$ & $0.20\pm0.09$ & $-2.97\pm0.15$ & $-1.1\pm2.3$ & $-3.5\pm2.9$\\
\hline
22.8 & $14.6\pm0.1$ & $12.8\pm0.1$ & $52.9\pm0.3$ & $-2.45\pm0.00$ & -- &-- & &$13.1\pm0.1$ & $10.4\pm0.1$ & $4.17\pm0.01$ & $-2.40\pm0.00$ & -- &--\\
33.0 & $6.9\pm0.1$ & $5.7\pm0.1$ & $18.4\pm0.3$ & $-2.48\pm0.00$ & $-2.0\pm0.0$ & $-2.2\pm0.0$ & & $6.2\pm0.1$ & $4.5\pm0.1$ & $1.72\pm0.01$ & $-2.40\pm0.00$ & $-2.0\pm0.1$ & $-2.3\pm0.0$\\
40.7 & $4.3\pm0.1$ & $3.5\pm0.1$ & $9.9\pm0.3$ & $-2.50\pm0.01$ & $-2.2\pm0.1$ & $-2.3\pm0.1$ & & $4.0\pm0.1$ & $2.7\pm0.1$ & $1.03\pm0.01$ & $-2.40\pm0.00$ & $-2.1\pm0.2$ & $-2.4\pm0.1$\\
\hline
22.8 & $8.2\pm0.5$ & $9.8\pm0.4$ & $9.2\pm2.4$ & $-2.88\pm0.07$ & -- &-- & &$8.0\pm0.5$ & $10.1\pm0.4$ & $0.58\pm0.12$ & $-3.27\pm0.09$ & -- &--\\
33.0 & $4.4\pm0.5$ & $3.7\pm0.4$ & $4.8\pm2.4$ & $-2.79\pm0.11$ & $-1.7\pm0.4$ & $-2.6\pm0.3$ & & $4.3\pm0.5$ & $3.8\pm0.4$ & $0.35\pm0.11$ & $-3.00\pm0.12$ & $-1.7\pm0.4$ & $-2.6\pm0.3$\\
40.7 & $3.3\pm0.5$ & $1.9\pm0.4$ & $4.0\pm2.4$ & $-2.70\pm0.13$ & $-1.4\pm1.0$ & $-3.0\pm1.2$ & & $3.1\pm0.5$ & $2.1\pm0.4$ & $0.25\pm0.11$ & $-2.90\pm0.16$ & $-1.4\pm1.0$ & $-2.9\pm1.0$\\
\hline
22.8 & $12.0\pm3.1$ & $5.8\pm1.9$ & $10.2\pm2.0$ & $-2.86\pm0.05$ & -- &-- & &$13.0\pm3.1$ & $6.1\pm1.9$ & $0.94\pm0.20$ & $-3.05\pm0.09$ & -- &--\\
33.0 & $7.9\pm3.0$ & $0.4\pm1.9$ & $6.2\pm1.9$ & $-2.73\pm0.07$ & $-1.1\pm1.3$ & $-7.1\pm12.0$ & & $8.9\pm3.0$ & $0.7\pm1.9$ & $0.32\pm0.19$ & $-3.03\pm0.23$ & $-1.0\pm1.1$ & $-5.7\pm6.9$\\
40.7 & $6.2\pm3.0$ & $-0.7\pm1.8$ & $5.2\pm1.9$ & $-2.64\pm0.08$ & $-1.1\pm2.9$ & -- & &$7.3\pm3.0$ & $-0.3\pm1.8$ & $0.15\pm0.19$ & $-3.09\pm0.45$ & $-1.0\pm2.5$ & --\\
\hline
22.8 & $1.5\pm2.9$ & $7.7\pm0.7$ & $6.8\pm0.9$ & $-2.96\pm0.03$ & -- &-- & &$2.2\pm2.9$ & $7.7\pm0.7$ & $0.97\pm0.08$ & $-3.04\pm0.04$ & -- &--\\
33.0 & $1.0\pm2.9$ & $2.0\pm0.7$ & $1.8\pm0.8$ & $-3.02\pm0.11$ & $-1.3\pm9.7$ & $-3.6\pm0.9$ & & $1.0\pm2.9$ & $2.0\pm0.7$ & $0.32\pm0.08$ & $-3.03\pm0.10$ & $-2.3\pm8.9$ & $-3.7\pm0.9$\\
40.7 & $-0.1\pm2.9$ & $0.7\pm0.6$ & $0.6\pm0.8$ & $-3.13\pm0.32$ & -- &$-5.1\pm4.7$ & & $-0.2\pm2.8$ & $0.6\pm0.6$ & $0.16\pm0.08$ & $-3.06\pm0.18$ & -- &$-5.4\pm5.1$\\
\hline
\end{tabular}
\label{tab:regioncoefficients}
\end{table*}

\bibliographystyle{mn2e}
\bibliography{references}
\bsp

\label{lastpage}

\end{document}